\documentclass{JHEP3}
\usepackage[latin1]{inputenc}
\usepackage{bbm,slashed}
\usepackage{graphicx,epsfig,}
\usepackage{amsmath,amsfonts,amssymb}
\newcommand{\psib}{\bar{\psi}}
\newcommand{\tbar}{\bar{\theta}}  
\newcommand{\Dbar}{\bar{D}}
\newcommand{\Zf}{\mathcal{Z}}
\newcommand{\N}{\mathcal{N}} 
\newcommand{\D}{\mathcal{D}}

\newcommand{\thetab}{\bar{\theta}}
\newcommand{\pa}{\partial}
\newcommand{\pat}{\partial_\theta}
\newcommand{\patb}{\partial_{\bar\theta}}
\newcommand{\eps}{\epsilon}
\newcommand{\epsb}{\bar\epsilon}
\newcommand{\veps}{\varepsilon}

\newcommand{\ha}{\frac{1}{2}}

\newcommand{\chib}{\bar\chi}

\newcommand{\FTild}{\mathcal{F}}

\DeclareMathOperator{\STr}{STr}
\DeclareMathOperator{\Tr}{Tr}

 \DeclareMathAlphabet{\boldmathe}{T1}{cmr}{bx}{it}

\newcommand{\cD}{{\mathcal D}}

\newcommand{\cW}{{\mathcal W}}

\newcommand{\mtxt}[1]{\quad\hbox{{#1}}\quad}

\def\P{\Phi}
\def\G{\Gamma}

\def\p{\partial}

\def\Gt{\Gamma^{(2)}}

\def\t{\theta}
\def\tb{\bar{\theta}}

\def\half{\frac{1}{2}}
\def\suint{d\theta d\thetab}
\def\intt{\int d\t d\tb\,}
\def\inttt{\int d\t d\tb\, d\t' d\tb'\,}
\def\tttt{d\t d\tb\, d\t' d\tb'\,}
\def\deltatt{\delta(\tb'-\tb)\delta(\t'-\t)}
\graphicspath{{plots/}}

\title{Flow Equation for Supersymmetric Quantum Mechanics} 
\author{Franziska Synatschke,  Georg Bergner, Holger Gies and Andreas
  Wipf\\
  Theoretisch-Physikalisches Institut, Friedrich-Schiller-Universit{\"a}t
Jena,
Max-Wien-Platz 1, D-07743 Jena, Germany\\
   E-mail: \email{franziska.synatschke@uni-jena.de, g.bergner@tpi.uni-jena.de,
gies@tpi.uni-jena.de, wipf@tpi.uni-jena.de}}

\abstract{
  We study supersymmetric quantum mechanics with the functional RG formulated
  in terms of an exact and manifestly off-shell supersymmetric flow equation
  for the effective action. We solve the flow equation nonperturbatively in a
  systematic super-covariant derivative expansion and concentrate on systems
  with unbroken supersymmetry. Already at next-to-leading order, the energy of
  the first excited state for convex potentials is accurately determined
  within a 1\% error for a wide range of couplings including deeply
  nonperturbative regimes. }
  
 \preprint{}
  

\begin{document}

\section{Introduction}

Supersymmetry is a key ingredient in the construction of models of fundamental
physics, since it provides for a salient possibility to combine internal
symmetries with the Poincare group. Even though distinguishing features of
supersymmetric systems can be understood within perturbation theory, many
important properties such as collective condensation phenomena often related to
symmetry breaking are inherently nonperturbative. If supersymmetry is realized
in nature, powerful and flexible nonperturbative tools will be needed to
investigate the underlying mechanisms of these strong-coupling phenomena. 

As supersymmetry does not only mix bosons and fermions but also involves
spacetime translations, lattice methods built on spacetime discretization
often go along with a partial loss of supersymmetry. The construction of
appropriate lattice formulations in addition to the challenge of dealing with
dynamical fermions is an ongoing effort
\cite{Feo:2004kx,Giedt:2006pd,Bergner:2007pu,Kastner:2008zc}.
These studies need to be
complemented by nonperturbative continuum methods preferably with manifest
supersymmetry. 

In recent years, the functional renormalization group (RG) has become such a
nonperturbative tool as has been demonstrated by many successful applications
ranging from critical phenomena, via fermionic systems and gauge theories even
to quantum gravity, see
\cite{Aoki:2000wm,Berges:2000ew,Litim:1998nf,Pawlowski:2005xe,Gies:2006wv,Sonoda:2007av}
for reviews. However, the number of applications to supersymmetric systems is
rather small. In this work, we formulate and test the functional RG for a
simple supersymmetric system, namely, supersymmetric quantum mechanics.

In fact, ordinary quantum mechanics has often been used for illustrating and
testing the nonperturbative capabilities of the functional RG, since the RG
flow equations are easily obtained and approximate solutions can directly be
compared to known exact results or high-precision numerics. In particular, the
study of ground- and excited-state energies with RG techniques has received a
great deal of interest in the last few years
\cite{Horikoshi:1998sw,Kapoyannis:2000sp,Zappala:2001nv,Weyrauch:2006,Gies:2006wv}.
 Whereas single-well potentials can be treated comparatively
easily even at extreme coupling, double-well potentials have turned
out to be more challenging, since the analytic RG flow equations have to build
up the non-analyticities from tunneling; the latter are usually described in
terms of instantons, being of topological nature.
 
In \cite{Horikoshi:1998sw}, Horikoshi et al.~study the quantum double well
using an expansion in powers of the field, \cite{Kapoyannis:2000sp} and
\cite{Weyrauch:2006,Weyrauch:2006aj} go beyond this approximation and solve
the RG flow in the so-called local-potential approximation for the effective
potential (i.e., leading-order derivative expansion). Within the propertime
RG, Zappal\`a \cite{Zappala:2001nv} also includes wave function
renormalization (i.e., next-to-leading-order derivative expansion), and finds 
good agreement for the mass gap. Particularly, this study convincingly
demonstrates that the functional RG automatically includes also fluctuations
of topological degrees of freedom without explicitly introducing them by
hand. 

Supersymmetric quantum mechanics was introduced by Witten \cite{Witten:1981nf}
as a toy model for spontaneous symmetry breaking.  The first to use
renormalization group methods for supersymmetric quantum mechanics were
Horikoshi et al.~\cite{Horikoshi:1998sw}. They investigated a broken
supersymmetric model with nonperturbative renormalization group methods and
calculated the nonvanishing ground-state energy and that of the first excited
state in a polynomial expansion of the effective potential. They found good 
agreement with the exact results for all cases where tunneling is not
important. This latter region has been covered in \cite{Weyrauch:2006aj}
within the propertime RG, where again the observation was made that a wave
function renormalization improves the results for the energy spectrum, i.e.,
helps including tunneling.

Both approaches use regulators that break supersymmetry which makes it
difficult to distinguish between explicit and spontaneous or dynamical
supersymmetry breaking. 
One possibility to solve this problem is the inclusion of symmetry breaking by the regulator into the symmetry relations as done in \cite{Bonini:1993sj,Ellwanger:1994iz}. 
In fact even the lattice discretisation can be viewed as a supersymmetry breaking regulator.
The corresponding modified symmetry relations are similar to the Ginsparg-Wilson-relation, introduced in \cite{Ginsparg:1981bj} and extended in \cite{Igarashi:2002ba}, and were established for supersymmetric models in \cite{Bergner:2008ws}. But so far a solution of these relations is possible only in some simple cases.
In this paper, we present an approach to flow
equations for supersymmetric quantum mechanics which maintains supersymmetry
manifestly on the level of the RG flow equation with the aid of an invariant
regulator. In contrast to \cite{Horikoshi:1998sw} and \cite{Weyrauch:2006aj},
we concentrate on a system with unbroken supersymmetry.

Our approach is similar to the works by Bonini and Vian
\cite{Vian:1998kv,Bonini:1998ec} where a supersymmetric regulator for the $4d$
Wess-Zumino model is presented. The functional RG has also been formulated for
supersymmetric Yang-Mills theory in \cite{Falkenberg:1998bg} employing the
superfield formalism; for applications, see also
\cite{Arnone:2004ey,Arnone:2004ek}. Very recently, Rosten has investigated
general theories of a scalar superfield including the Wess-Zumino model with
the aid of a Polchinski-type of RG equation with elegant applications in the
context of non-renormalization theorems \cite{Rosten:2008ih}. A construction
of a Wilsonian effective action for the Wess-Zumino model by perturbatively
iterating the functional RG has been performed in \cite{Sonoda:2008dz}. 

The paper is organized as follows: in Sect.~\ref{sec:susyQM}, we briefly
recall the basics of Euclidean supersymmetric quantum mechanics, also
introducing our notation. In Sect.~\ref{sec:flowEqOfShell} we derive the flow
equation for the superpotential and introduce a general class of supersymmetric
regulator functions. In Sect.~\ref{sec:flowDiffReg} we discuss the flow equation
for the superpotential for different regulators. In Sect.~\ref{sec:WFR} we introduce
wave function renormalization and in Sect.~\ref{sec:Compare} we compare our
results with exactly known results.

\section{Euclidean supersymmetric quantum mechanics}
\label{sec:susyQM}
For our study of supersymmetric quantum mechanical RG flows, we employ the
superfield formalism to maintain supersymmetry manifestly.  The Euclidean
superfield has the expansion
\begin{align}
\Phi=\phi +\thetab\psi +\psib \theta+\thetab\theta F\label{susy1}
\end{align}
with anticommuting parameters $\theta,\thetab$.
Supersymmetric interaction terms are obtained as $D$-term of
\begin{align}
W(\Phi)=W(\phi)+\big(\thetab\psi+\psib\theta\big)W'(\phi)+
\thetab\theta\big(F W'(\phi)-W''(\phi)\psib\psi\big),\label{susy4}
\end{align} where  the
superpotential $W(\Phi)$ is a polynomial in $\Phi$, and $W(\phi)$ 
denotes the same polynomial evaluated at the scalar field $\phi$. 
The nilpotent supercharges
$
Q=i\patb+\theta\pa_\tau$ {and} $\bar Q=i\pat+\thetab\pa_\tau$
anticommute into the generator of (Euclidean) time-translations,
$Q\bar Q+\bar QQ=2i\pa_\tau$.
Supersymmetry variations are generated by
$
\delta_\eps=\epsb Q-\eps\bar Q$,
such that the variation of the superfield takes the form
\begin{align}
\delta_\eps\Phi
=\epsb\big(i\psi+i\theta F+\theta\dot\phi+\theta\thetab\dot\psi\big)
-\big(i\psib+i\thetab F
-\thetab\dot\phi+\thetab\theta\dot{\psib}\big)\eps,\label{susy9} 
\end{align}
from which we read off the transformation rules for the component
fields,
\begin{align}
\delta\phi=i\epsb\psi-i\psib\eps\,, \qquad
\delta\psi=(\dot\phi-iF)\eps\,,\qquad
\delta\psib=\epsb(\dot\phi+iF),\qquad
\delta F=-\epsb\dot\psi-\dot{\psib}\eps\,.
\label{susy11}
\end{align}
The super-covariant derivatives
$D=i\patb-\theta\pa_\tau$ and $\bar D=i\pat-\thetab\pa_\tau$
fulfill similar anticommutation relations as the supercharges,
\begin{align}
\{D,D\}=\{\bar D,\bar D\}=0\mtxt{and}\{D,\bar D\}=-2i\pa_\tau\,.\label{susy13}
\end{align}
They commute with $\pa_\tau$ and anticommute with the 
supercharges.
The integration over the anticommuting variables extracts the
D-term of a superfield
\begin{align}
\int d\theta d\thetab \Phi\equiv\Phi\vert_{\thetab\theta}\,.\label{susy17}
\end{align}
From this, we obtain the invariant action in the superfield formalism:
\begin{align}
   	S[\phi,F,{\psib},\psi]&=
    \int d\tau \suint
  	\left[\frac12\Phi
	K\Phi+i\cdot W(\Phi)\right]\nonumber\\
& =
  \int d\tau
  	\left[\frac{1}{2}\dot\phi^2-i\psib\dot\psi
  	+iFW'(\phi)-i\psib W''(\phi)\psi+\frac12 F^2\right],\label{susy19}
  	\end{align}
with kinetic operator $K=\ha(D\Dbar -\Dbar D)$.
Eliminating the auxiliary field $F$, we obtain the
\emph{on-shell} action
\begin{align}
	S_{\rm on}[\phi,\psi,\psib]
	=\int d\tau\left[\frac{1}{2}\dot\phi^2-i\psib\dot\psi 
	+\frac12\big(W'{}(\phi)\big)^2-iW''(\phi)\psib\psi
	\right].\label{susy21}
\end{align}
It contains the bosonic potential 
 $V(\phi)=\frac12\big(W'(\phi)\big)^2$ and a Yukawa term. 
In this paper, we consider models with unbroken supersymmetry. 
They have vanishing ground state energy $E_0=0$  and are realized for
 superpotentials whose highest power is \emph{even}. On the microscopic scale, we will
 focus on quartic superpotentials
\begin{align}
W(\phi)=e\phi+\frac m2\phi^2+\frac g3\phi^3+\frac a4\phi^4\,,\label{susy23}
\end{align}
as the defining starting point of the  interactions of our quantum mechanical system
before fluctuations are taken into account.

\section{Flow equation in the off-shell formulation}
\label{sec:flowEqOfShell}

\subsection{Flow equation for the effective action}

The functional RG can be formulated in terms of a flow equation for the {\em
  effective average action} $\Gamma_k$ \cite{Wetterich:1992yh}. This is a
scale-dependent action functional which interpolates between the microscopic
or classical action $S$ and the full quantum effective action $\Gamma$, being
the generating functional for 1PI Green's functions. The interpolation scale
$k$ denotes an infrared $IR$ regulator scale which suppresses all fluctuations
with momenta smaller than $k$. For $k\to\Lambda$ with $\Lambda$ denoting the
microscopic scale, no fluctuations are included such that $\Gamma_{k\to
  \Lambda}\to S$. For $k\to0$, all fluctuations are taken into account and we
arrive at $\Gamma_{k\to0}\to \Gamma$, i.e., the full solution of the quantum
theory. The effective average action can be determined from the Wetterich
equation \cite{Wetterich:1992yh}
\begin{equation}
 \partial_k\Gamma_k=
 \frac12 \STr\left\{\left[\Gamma_k^{(2)}+ R_k\right]^{-1}\partial_k  R_k\right\}
\label{eq:wetterich}
\end{equation}
which defines an RG flow trajectory in the space of action functionals with
the classical action serving as initial condition. Here, $\Gamma^{(2)}$
denotes the second functional derivative with respect to the dynamical
fields, 
\begin{equation}
\left(\Gamma_k^{(2)}\right)_{ab}=\frac{\overrightarrow{\delta}}{\delta\Psi_a}
\Gamma_k\frac {\overleftarrow{\delta}}{\delta\Psi_b}\,,\label{eq:G2}
\end{equation}
where the indices $a,b$ in the general case summarize field components,
internal and Lorentz indices, as well as spacetime or momentum coordinates. In
the present case, we have $\Psi^{\text T}=(\phi,F,\psi,\bar\psi)$. (Note that
$\Psi$ is not a superfield, but merely a collection of fields.) The supertrace
in eq. \eqref{eq:wetterich} as well as the pattern of functional
differentiation in eq. \eqref{eq:G2} takes care of the minus signs from
Grassmann-valued variables. The regulator function $R_k$ guarantees the IR
suppression of modes below $k$, the shape of which is to some extent
arbitrary; examples will be given below. Different $R_k$ correspond to
different RG trajectories manifesting the RG scheme dependence, but the end
point $\Gamma_{k\to 0}\to \Gamma$ remains invariant. 

The flow equation \eqref{eq:wetterich} is an exact equation, involving the
regularized exact propagator $G_k\equiv (\Gamma_k^{(2)}+R_k)^{-1}$, and has a
one-loop structure. It can be viewed as the differential counterpart of a
functional integral, or path integral in quantum mechanics. Its perturbative
expansion yields full standard perturbation theory, but also nonperturbative
systematic expansion schemes can be devised. In the present work, we use a
derivative expansion of the effective action in powers of the covariant
derivative in the off-shell formulation. This expansion is systematic in the
sense that all possible operators can uniquely be classified, and it is
consistent, since dropping higher-order terms leads to a closed set of
equations.  Most importantly, a truncation of such an expansion preserves
supersymmetry.  In this work, the derivative expansion of supersymmetric
quantum mechanics will be worked out to next-to-leading-order. For simplicity,
let us here begin with the leading order, corresponding to the local-potential
approximation for the superpotential; to this order, the truncated effective
action reads
\begin{align}
  \Gamma_k[\phi,F,{\psib},\psi]&= \int d\tau\suint \left[\frac12 \Phi
    K\Phi+i\cdot W_k(\Phi)\right]\nonumber\\
 &= \int d\tau
  \left[\frac{1}{2}\dot\phi^2-i\psib\dot\psi +\frac12
    F^2+iFW'_k(\phi)-iW''_k(\phi)\psib\psi\right].\label{flow1}
 \end{align}
The prime always denotes the derivative with respect to the bosonic field
$\phi$.  In the following we will derive flow equations for the superpotential
$W_k(\phi)$. The next order which includes a wave function renormalization
will be considered later on.

Let us finally mention that the effective action is particularly convenient
for extracting physical quantities: the effective action $\Gamma=\Gamma_{k=0}$
evaluated on the solution of its quantum equation of motion yields the ground
state energy, which is zero if supersymmetry is unbroken. Since $\Gamma$ is
the generating functional of 1PI Green's functions, it provides access to all
correlators and corresponding quantities. For instance, the location $p^2$ of
the pole of the propagator $\Gamma^{(2)}(p^2)=0$ is a measure for the energy
of the first excited state in supersymmetric quantum mechanics (corresponding
to particle masses in quantum field theory). In the derivative expansion, this
excited-state energy can directly be related to properties of the
superpotential, see below. An alternative to the effective-action flow would
be the flow of the Wilson action $S_k$ which has the advantage of being
regulator-independent at leading-order in the derivative expansion
\cite{Morris:2005ck}, but can suffer from numerical instabilities within
truncations \cite{Pawlowski:2005xe}.

\subsection{Supersymmetric regulators}

For a supersymmetric initial condition and truncation, the flow and the
resulting effective action is supersymmetric provided the regulator does not
break the symmetry. When deriving the flow equation \eqref{eq:wetterich} from
the functional integral, the regularization is introduced by means of an
additional action contribution $\Delta S_k$, such that $R_k= \Delta
S_k^{(2)}$. The action principle therefore guarantees a supersymmetric
regularization, as long as $\Delta S_k$ is invariant.  Indeed, an off-shell
supersymmetric cutoff action can be written in terms of superfields and its
covariant derivatives:
\begin{align}
\Delta S_k =\frac12  \int d\tau\; \Phi R_k(D,\Dbar)
\Phi|_{\tbar\theta}\,.\label{reg1} 
\end{align}
Since $D$ and $\bar D$ satisfy the anticommutation relations \eqref{susy13}
the regulator can be written as 
\begin{align}
R_k(D,\Dbar)=
	ir_1(-\partial_\tau^2,k)+r_2(-\partial_\tau^2,k)K\,,\qquad
K=\ha(D\Dbar-\Dbar D). \label{reg3}
\end{align}
The factor $i$ in front of $r_1$ is chosen for convenience such that the
corresponding cutoff action matches the mass term. Similarly $r_2$ is chosen
such that its cutoff action matches the kinetic term. Both functions are
functions of $-\partial_\tau^2$, i.e., of $p^2$ in momentum space. For this
general class of regulators, the cutoff actions read
\begin{align}
 \Delta S_k=& \ha\int d\tau\suint\; \Phi\left( ir_1+r_2K \right)\Phi 
=\ha \int \frac{dp}{2\pi}\Psi^T(-p)  R_k(p)\Psi(p)\,,
\label{reg5}
\end{align}
where $\Psi^T=(\phi,F,\psi,\bar\psi)$. 
The quadratic form $R_k(p)$ is block-diagonal,
\begin{align}
 R_k=\begin{pmatrix} R_k^{\rm B}&0\\ 0& R_k^{\rm F}\end{pmatrix}
\mtxt{with blocks}
 R_k^{\rm B}=\begin{pmatrix}p^2r_2&ir_1\\ ir_1&r_2\end{pmatrix},\quad
 R_k^{\rm F}=\begin{pmatrix}0&pr_2+ir_1\\ pr_2-ir_1&0\end{pmatrix},
\label{reg7}
\end{align}
and hence does not mix bosonic and fermionic degrees of freedom. Three
properties of the regulator $R_k(p)$ are essential: (i) $R_k(p)|_{p^2/k^2\to
  0} >0$ in order to implement an IR regularization, (ii) $R_k(p)|_{k^2/p^2\to
  0} =0$, implying that the regulator vanishes for vanishing $k$, (iii)
$R_k(p)|_{k\to\Lambda\to\infty}\to \infty$ which helps fixing the theory with
the classical action in the UV.

For manifestly supersymmetric cutoff actions $\Delta S_k$, supersymmetry
relates the regulators of bosonic fields to that of the fermionic field. This
puts further constraints on the admitted cutoff functions in a supersymmetric
theory, as can be seen from the following example. In view of the regulator
structure in eq.~\eqref{reg7}, one may be tempted to set $r_1=0$. A natural
choice for the regulator functions would then be such that the bosonic
component $\sim p^2r_2$ induces a gap for IR modes, e.g., $r_2(p^2/k^2) \sim
k^2/p^2$ such that $p^2 r_2 \sim k^2$.  Supersymmetry implies to the
regulator $pr_2$ for the fermions and to the regulator $r_2$ for the auxiliary
field, both of which diverge in the IR for this choice. 
Even though regulators of this type are perfectly legitimate in the full flow
equation, they lead to artificial IR divergencies at
higher order in the derivative expansion, e.g., for a wave function
renormalization. This problem can be avoided by a softer IR behavior of $r_2$
and including a suitable nonvanishing $r_1$.

\subsection{Regularized on-shell action}
The equation of motion for the auxillary field in the presence of  a cutoff is
\begin{align}
	F=-\frac{i}{h}\cW'\,,\qquad \cW'(\phi)=W_k'(\phi)+r_1\phi\,,\qquad
h=1+r_2\,,\label{reg9}
\end{align}
where, for convenience, we have introduced the
function $h(p)$ and the shifted superpotential $\cW$ containing the cutoff 
functions $r_2$ and $r_1$.
The regularized non-local on-shell action becomes
\begin{align}
	S_{\rm on}=\int d\tau \left[\frac12\dot\phi h\dot\phi-i\psib h\dot\psi
	-i\psib \cW''(\phi)\psi+\frac12 \cW'(\phi)\frac{1}{h}
	\cW'(\phi)\right].\label{reg11} 
\end{align}
It is invariant under the following \emph{deformed supersymmetry} transformations
\begin{align}
\delta\phi=i\epsb\psi-i\psib\eps\,,\qquad
\delta\psi=\left(\dot\phi-\frac{1}{h}\cW'(\phi)\right)\eps\,,\qquad
\delta\psib=\epsb \left(\dot \phi+\frac{1}{h}\cW'(\phi)\right).\label{reg13}
\end{align}
These non-local transformations close on infinitesimal
time translations,
\begin{align}
(\delta_{\eps_2}\delta_{\eps_1}-\delta_{\eps_1}\delta_{\eps_2})(\hbox{field})
=2i(\epsb_1\eps_2-\epsb_2\eps_1)\pa_\tau(\hbox{field})\,,\label{reg15}
\end{align}
provided the fermionic field satisfies the deformed Dirac equation
$
h\dot\psi+\cW''(\phi)\psi=0
$. With \eqref{reg11} we have
constructed a \emph{regularized} (nonlocal) on-shell action which is
invariant under deformed supersymmetry transformations.

Nevertheless, we would like to stress that the off-shell formulation is
crucial for the construction of an invariant flow equation with one-loop
structure. As the on-shell supersymmetry transformations act nonlinearly on
the fields, the resulting cutoff action is not quadratic in the fields. Even
though an on-shell supersymmetric flow can straightforwardly be constructed
from eq.~\eqref{reg11}, the resulting flow involves higher-loop terms and thus
is much more difficult to deal with.

\subsection{Flow equation}
Returning to the off-shell formulation and using the block-diagonal structure
of the regulator \eqref{reg7}, the flow equation for the 
effective action $\Gamma_k[\phi,F,\psib,\psi]$ written in
component fields reads
\begin{align}
  \partial_k\Gamma_k=
 \frac12 \STr\left\{\left[\Gamma_k^{(2)}+ R_k\right]^{-1}\partial_k  R_k\right\}
=\ha\Tr \left({\pa_kR_k}\,{G_k}\right)_{BB}
 -\ha\Tr \left({\pa_k R_k}\,{G_k}\right)_{FF}\,,\label{flow3}
 \end{align}
where we have introduced the regularized full Green's function
or propagator $G_k=(\Gamma_k^{(2)}+R_k)^{-1}$. 
Upon insertion of the truncation \eqref{flow1} into eq.~\eqref{flow3}, we need
to project only onto the flow of the superpotential $W_k$. It can be done
by extracting the flow of either the term linear in $F$ or the term
proportional to $\bar\psi\psi$, cf. eq.~\eqref{flow1}. This is a direct
consequence of the manifest supersymmetry of this approach.  As an illustration 
of this fact, we do it both ways. For the projection, it suffices to
consider constant fields, 
such that an expansion of the inverse Green's function in terms of the constant
anticommuting spinors $\psi,\psib$ yields
\begin{align}
G_k^{-1}=\Gamma^{(2)}_k+ R_k\equiv G_{0,k}^{-1}+\psib M_1+M_2\psi+\psib M_3\psi\,.
\label{flow7}
\end{align}
The propagator itself reads
\begin{align}
G_k=&\,G_{0,k}-G_{0,k}(\psib M_1 +M_2\psi)G_{0,k} \nonumber\\
&+G_{0,k}\left(M_1 G_{0,k} M_2-M_2G_{0,k}
M_1-M_3\right)G_{0,k}\psib\psi\,.\label{flow9} 
\end{align}
 To proceed we use the block notation,
\begin{align}
N=\begin{pmatrix}N_{BB}&N_{BF}\\ N_{FB}&N_{FF}\end{pmatrix}.\label{flow11}
  \end{align}
The nonvanishing blocks of the operators in the expansion \eqref{flow7} have
the form
\begin{equation}
\begin{aligned}  
(G_{0,k}^{-1})_{BB}&=\begin{pmatrix}
hp^2+iF\cW^{(3)}&i\cW''\\ i\cW''&h\end{pmatrix} 
,\qquad (G_{0,k}^{-1})_{FF}=
\begin{pmatrix}0&hp+i\cW''\\ hp-i\cW''&0\end{pmatrix},\\
M_{1FB}&=-M_{1BF}=\begin{pmatrix}i\cW^{(3)} &0\cr 0&0\end{pmatrix},\qquad
M_{2BF}=-M_{2FB}^T=\begin{pmatrix}0&i\cW^{(3)}\cr 0&0\end{pmatrix},\\
M_{3BB}&=\begin{pmatrix}-i\cW^{(4)}&0\\ 0&0\end{pmatrix}.
\label{flow13}
\end{aligned}
\end{equation}

To calculate the full propagator $G_k$ we must invert $G_{0,k}^{-1}$.
The inverse of $G_{0,k}^{-1}$ is block diagonal, and the diagonal blocks read
for constant fields 
\begin{align}
(G_{0,k})_{BB}=\frac{1}{\Delta_B}\begin{pmatrix}h&-i\cW''\\ -i\cW''&hp^2+iF\cW^{(3)}\end{pmatrix}\mtxt{and}
(G_{0,k})_{FF}=\frac{1}{\Delta_F}\,(G_{0,k}^{-1})_{FF}\label{flow17}
\end{align}
with determinantal factors 
\begin{align}
\Delta_{\rm F}=h^2p^2+(\cW'')^2\mtxt{and}
\Delta_{\rm B}=\Delta_{\rm F}+ihF\cW^{(3)}\,.\label{flow19}
\end{align}
Since the regulator $R_k$ is block-diagonal, see \eqref{reg7}, only the diagonal
blocks of the dressed propagator enter the flow equation \eqref{flow3}. 
These blocks can be calculated with the help of \eqref{flow9}. Inserting 
the regulator \eqref{reg7} finally yields
\begin{align}
\hbox{Str}\left(\pa_k R_k \,G_k\right)=\int d\tau\left(H_0+H_3\psib\psi\right)
\label{flow21}
\end{align}
with $\phi$ and $F$-dependent coefficient functions
\begin{align}
H_0&=
-iF\cW^{(3)}\int\frac{dp}{2\pi}\,\frac{\pa_k r_2 (h^2p^2-\cW''^{\,2})+2h\pa_k
r_1\cW''}{\Delta_B\Delta_F}\label{flow23} 
\intertext{and}
H_3&=i\int\frac{dp}{2\pi}\nonumber
\left(\Delta_F\cW^{(4)}-2(\cW^{(3)})^2\cW''\right)
\frac{\pa_k r_2 (h^2p^2-\cW''^2)+2h\pa_k r_1 \cW''}{\Delta_B^2\Delta_F}\\
&\quad+2i\int \frac{dp}{2\pi}h(\cW^{(3)})^2\;
\frac{\pa_k r_1(h^2p^2-\cW''^2)
-2hp^2\pa_k r_2\cW''}{\Delta_B\Delta_F^2}\,.
\label{flow25}
\end{align}
The flow equation \eqref{flow3} relates the supertrace \eqref{flow21} 
to the variation of the effective action \eqref{flow1}. To project onto the 
flow for the superpotential, we differentiate the flow equation
with respect to $F$ and afterwards set $F=\psi=\psib=0$. This yields
\begin{align}
\pa_k W_k'=-\frac{i}{2} \frac{\pa \Gamma_0}{\pa F}\Big\vert_{F=0}
=-\frac{\cW^{(3)}}{2}\int\frac{dp}{2\pi}\,
\frac{\pa_k r_2 (h^2p^2-\cW''^{\,2})+2h\pa_k
r_1\cW''}{\Delta^2_B}\,.\label{flow27} 
\end{align}
Integrating with respect to $\phi$ (and dropping the irrelevant constant of
integration) finally yields the flow equation for the superpotential
\begin{align}
\pa_k W_k(\phi)=\ha \int \frac{dp}{2\pi}
	\frac{h\pa_k r_1-\pa_kr_2\cW''(\phi)}
		{h^2p^2+\cW''(\phi)^2}\,,\label{flow29}
\end{align}
where we recall the abbreviations $h=1+r_2$ and $\cW''=r_1+W_k''$.
This flow equation for the superpotential is one of the central results
of our work. From the solution of \eqref{flow29}, we can calculate
the effective potential $V_k$ by eliminating the auxiliary field in the
effective action. In passing, we note that a quicker way to obtain the flow
equation makes use of the superspace formulation, and an efficient approach is
summarized in appendix \ref{FlowSuperspace}.

The flow equation \eqref{flow29} can alternatively be
obtained by projecting the flow of the effective action \eqref{flow3}
onto the coefficient of $\psib\psi$. This way one obtains
\begin{align}
\pa_k W_k''=\ha H_3\big\vert_{F=0}\,.\label{flow31}
\end{align}
The two projection formulas \eqref{flow27} and \eqref{flow31} indeed give
rise to identical flows, since
\begin{equation}
\frac{\pa^2 H_0}{\pa\phi\pa F}\vert_{F=0}=i H_3\vert_{F=0}\,.
\end{equation}
This identity illustrates the fact that our flow equation is manifestly
supersymmetric. 
 
\section{Flow of the superpotential  for different regulators}
\label{sec:flowDiffReg}

The regulator in the flow equation not only suppresses IR modes, but also
guarantees UV regularization due to the operator insertion $\partial_k R_k$
for $R_k$ decreasing with momentum. This renders the flow local in momentum
space, enhancing also the numerical stability. In quantum mechanics, this
property is less important, since quantum mechanics is UV finite. This allows
to choose less UV-restrictive regulators for which the momentum integral in
eq.~\eqref{flow29} can be carried out analytically. 

Indeed, as long as no diagrams with closed $F$ loops contribute to the
truncation, the regulator $r_2$ can be dropped completely, as $r_1$ is
sufficient to regularize all diagrams with at least one $\phi$ or $\psi$ line,
as is clear from the structure of the regulator \eqref{reg7}.  Then the flow
equation \eqref{flow29} simplifies to
\begin{align}
\pa_k W_k(\phi)=\ha \int_{-\infty}^\infty \frac{dp}{2\pi}
\frac{\pa_k r_1}{p^2+(r_1+W''_k(\phi))^2}\,.
\label{reflow1}
\end{align}
We verify in appendix \ref{appB}, that this regulator choice is sufficient for
guaranteeing that the microscopic action is the correct starting point of the
flow without closed $F$ loops.  Incidentally, setting $r_1=0$ and using $r_2$
as a regulator alone in the flow equations would lead to artificial
divergencies for the wave function renormalization, as mentioned above.  Next,
we will discuss and compare different regulators. In principle, the choice of
the regulator can be optimized in order to minimize truncation
artifacts. However, due to the mixing between momentum- and field-dependencies
in the denominator of eq. \eqref{reflow1} $\sim r_1(p^2) W_k''(\phi)$, simple
optimization strategies for bosonic systems \cite{Litim:2000ci,Litim:2001up}
do not apply and full functional optimization would be required
\cite{Pawlowski:2005xe}. However, since we are not aiming for high-precision
calculations, our regulator choice will be guided by simplicity.

\subsection{The Callan-Symanzik regulator}
First, we consider a simple Callan-Symanzik regulator $r_1(p^2,k)=k$ for which
eq. \eqref{reflow1} reduces to the simple flow equation
\begin{align}
\partial_k W_k(\phi) =\frac14\cdot\frac{1}{k+W_k''(\phi)}\,.\label{reflow3}
\end{align}
We will discuss and compare various approaches to solve this flow equation for
different parameters and, in particular, for non-convex classical
superpotentials.

\subsubsection{Polynomial expansion}
\label{sect:polyexp}

For a polynomial approximation, one may expand the superpotential $W_k(\phi)$ in eq.
\eqref{reflow3} in powers of the bosonic field $\phi$,
\begin{align}
W_k(\phi)=\sum_{n} \frac{a_n(k)}{n}\,\phi^n\mtxt{with}
W_{k\to\Lambda}(\phi)=W_{\rm cl}(\phi)=e\phi+\frac{m}{2}\phi^2+\frac{g}{3}\phi^3+\frac{a}{4}\phi^4\,.
\label{polex1}
\end{align}
Then also the right hand side of the flow equation can be expanded similarly. A
comparison of coefficients leads to a system of coupled ordinary differential
equations for the coefficients 
$a_n(k)$. Terminating the expansions  on both sides at order $N$ and setting $a_{n>N}\to0$
the system becomes closed and can be solved
numerically. At the cutoff $k=\Lambda$, the non-vanishing coefficients are
$(a_1,a_2,a_3,a_4)=(e,m,g,a)$.  Note that for $g^2>3ma$ the classical
superpotential becomes non-convex.

Indeed, such an expansion about $\phi=0$ is not adjusted to the flow, as the
largest contribution to the flow equation arises from field values which
minimize  $W_k''$. An expansion of eq. \eqref{polex1} about the minimum of $W_k''$,
\begin{align}
W_k(\phi)=\sum_{n=1}^N \frac{\tilde a_n(k)}{n} \big(\phi-\phi_0(k)\big)^n\,,\qquad
W'''_k(\phi_0)=
2\tilde a_3=0\,,\label{polex3}
\end{align}
thus has a much better convergence behavior.  At the cutoff, the initial
conditions are provided by the nonvanishing parameters $(\tilde a_1,\tilde
a_2,\tilde a_4,\phi_0)$ which can directly be linked with $(e,m,g,a)$ given
above. Most importantly, $W_\Lambda''= \tilde a_2+3\tilde a_4(\phi-\phi_0)^2$
is an even function of $\phi-\phi_0$. Thus, the flow is also even, implying
that $W''_k$ stays even at all scales and all coefficients $\tilde a_n(k)$
vanish for odd $n\geq 3$. From the $(\phi-\phi_0)^3$ coefficient of the flow,
we find $(k+\tilde a_2)^2\pa_k\phi_0=\tilde a_5/\tilde a_4=0$ which states
that $\phi_0$ is \emph{scale-invariant}. The same is true for $\tilde a_1$,
since $\pa_k \tilde a_1=0$. The differential equations for the nontrivial even
coefficients of the truncated system up to order $N=10$ read
\begin{eqnarray*}
\pa_k \tilde a_2&=&-\frac{3}{2}\frac{\tilde a_4}{P^2}\,,\qquad P=k+\tilde a_2\\
\pa_k \tilde a_4&=&\hskip5mm\frac{9\tilde a_4^2-5\tilde a_6P}{P^3}\\
\pa_k \tilde a_6&=&-\frac{3}{2}\frac{27\tilde a_4^3-30\tilde a_4\tilde a_6P+7\tilde a_8P^2}{P^4}\\
\pa_k \tilde a_8&=& \hskip3mm 2\,\frac{81\tilde a_4^4-135 \tilde a_4^2\tilde a_6 P+(25\tilde a_6^2+42\tilde a_4\tilde a_8)P^2-9\tilde a_{10}P^3
}{P^5}\\
\pa_k \tilde a_{10}&=&-\frac{5}{2}\frac{243 \tilde a_4^5-540 \tilde a_6\tilde a_4^3P
+(189 \tilde a_8\tilde a_4^2+225 \tilde a_6^2\tilde a_4)P^2-70\tilde a_8\tilde a_6P^3-54\tilde a_{10}\tilde a_{4} P^4}{P^6} \,.
\end{eqnarray*}
The energy $E_1$ of the first excited state is determined by the curvature of
the effective potential $V_k=\frac12 (W'_k)^2$ at its minimum $\phi_{\rm
  min}$; note that $\phi_{\text{min}}$ is generically not equal to
$\phi_0$. At the minimum, $W'$ vanishes, such that $E_1=W''(\phi_{\text{min}})$. Table
\ref{tab:FirstExcitedState} contains the gap energy $E_1$ for classical
superpotentials with parameters $e=m=a=1$ and different values of $g$.
\TABLE{
\begin{tabular}{c|cccccccccc}
g&0.0&0.2&0.4&0.6&0.8&1.0&1.2&1.4&1.6&1.8\\ \hline
$\phi^4$ &2.008&1.960&1.895&1.815&1.722&1.615&1.497&1.371&1.237&1.097\\
$\phi^6$ &2.205&2.140&2.064&1.980&1.889&1.794&1.699&1.608&1.530&1.472\\
$\phi^8$ &2.214&2.146&2.070&1.987&1.898&1.808&1.721&1.646&1.596&1.590\\
$\phi^{10}$&2.201&2.135&2.060&1.977&1.888&1.798&1.711&1.638&1.595&1.612\\ \hline
PDE&2.203&2.137&2.062&1.979&1.890&1.798&1.710&1.633&1.584&1.590\\ \hline
exact&2.022&1.970&1.905&1.827&1.738&1.639&1.534&1.426&1.323&1.235
\end{tabular}
\caption{Energy $E_1$ of the first excited state calculated in different
  orders of the polynomial approximation with the Callan-Symanzik regulator
  for $e=m=a=1$.  For comparison, also the results obtained from the solution to the
  partial differential equation 
  \eqref{reflow3}
  (PDE) and the exact values
  from numerically diagonalizing the Hamiltonian are
  given.
  \label{tab:FirstExcitedState}}
}
For $g^2>3$ the initial superpotential becomes non-convex. In addition, the
minimum $\phi_{\text{min}}$ moves away from the expansion point $\phi_0$, in
principle signaling the break down of the polynomial approximation which can
be expected to hold only near $\phi_0$.  Nevertheless, the values $E_1(g)$
obtained for the polynomial approximations of orders $4,6,8$ and $10$ converge
to values obtained by solving the full partial differential equation
\eqref{reflow3}. We conclude that the polynomial expansion as an approximation
to the full solution to leading-order derivative expansion works
satisfactorily for the energy $E_1$ at these coupling values.  However, as the
$\sim 10\%$ difference to the exact gap energies shows, the leading-order
derivative expansion itself gives acceptable but not very precise
results. This should be compared to the analogous flow-equation approximation
for non-supersymmetric quantum mechanics which yields an error below the
percent level even at strong coupling.

One important difference is that we have a flow equation for the
superpotential and not for the effective potential itself. As a consequence,
the flow equation tends to make the superpotential convex but not necessarily
the effective potential.  Figure \ref{fig:polynomial} shows the flow of the
effective potential $V_k$ in the polynomial approximation \eqref{polex3} with
$N=6$ for a convex and non-convex $W_{\rm cl}$.

 \FIGURE{
\includegraphics[width=.49\textwidth]{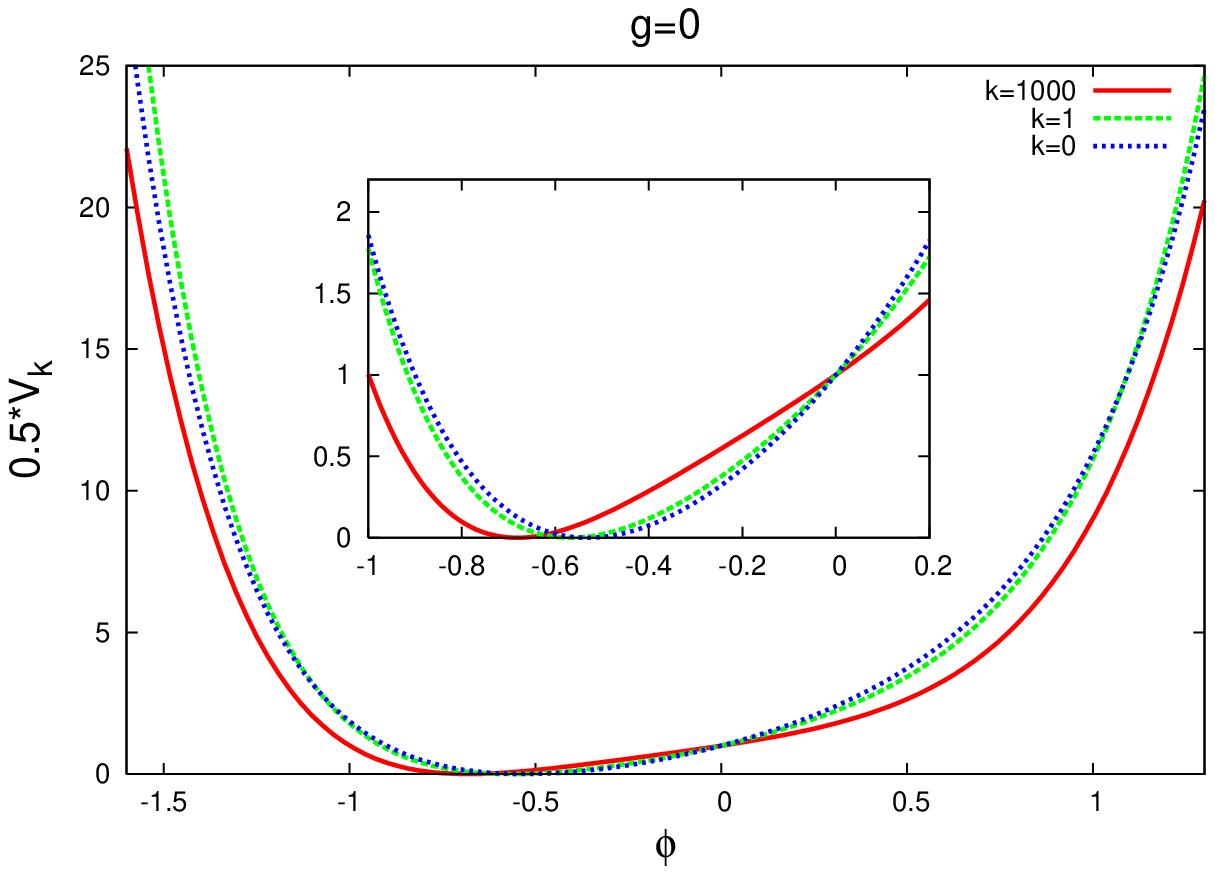}
\hfill
\includegraphics[width=.49\textwidth]{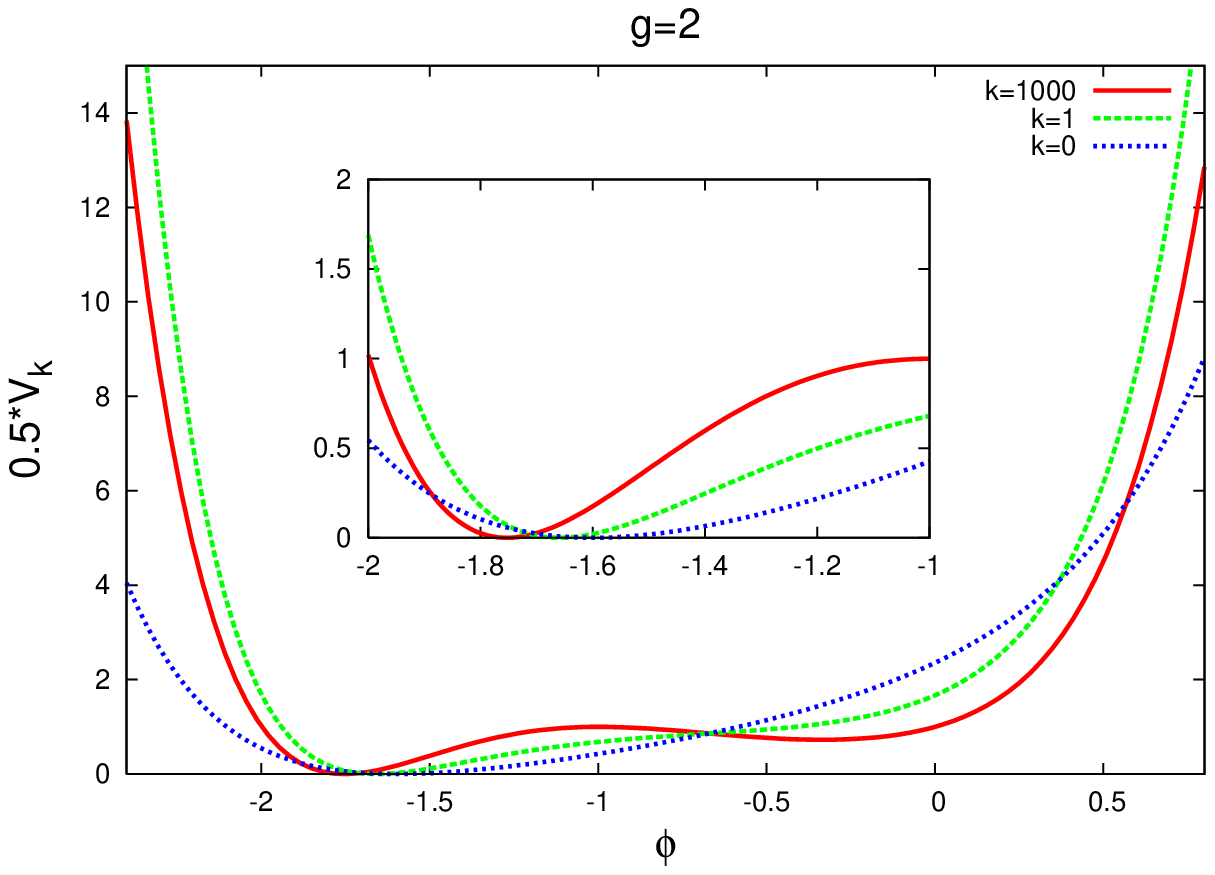}
\caption{The effective potential $V_k$ in the
\emph{polynomial approximation} for $W_{\rm cl}'(\phi)=1+\phi+g\phi^2+\phi^3$. The 
left panel shows the potential for $g=0$ and the right panel for
$g=2$.\label{fig:polynomial}}
}

\subsubsection{Partial differential equation}
It is known from the study of non-supersymmetric systems that the polynomial
approximation fails for nonconvex potentials
\cite{Kapoyannis:2000sp,Weyrauch:2006aj}. The latter require a solution of the
full partial differential equation \eqref{reflow3}, which we did with
\texttt{NDSolve} of
\textsc{Mathematica}. In practice, we have chosen $\phi$ in the range of
$\phi\in(-200,200)$ and kept the potential at its classical values on the
boundary of this range. The results for three different scales are depicted in
Figure \ref{fig:EffPot}.
\FIGURE{
\includegraphics[width=.49\textwidth]{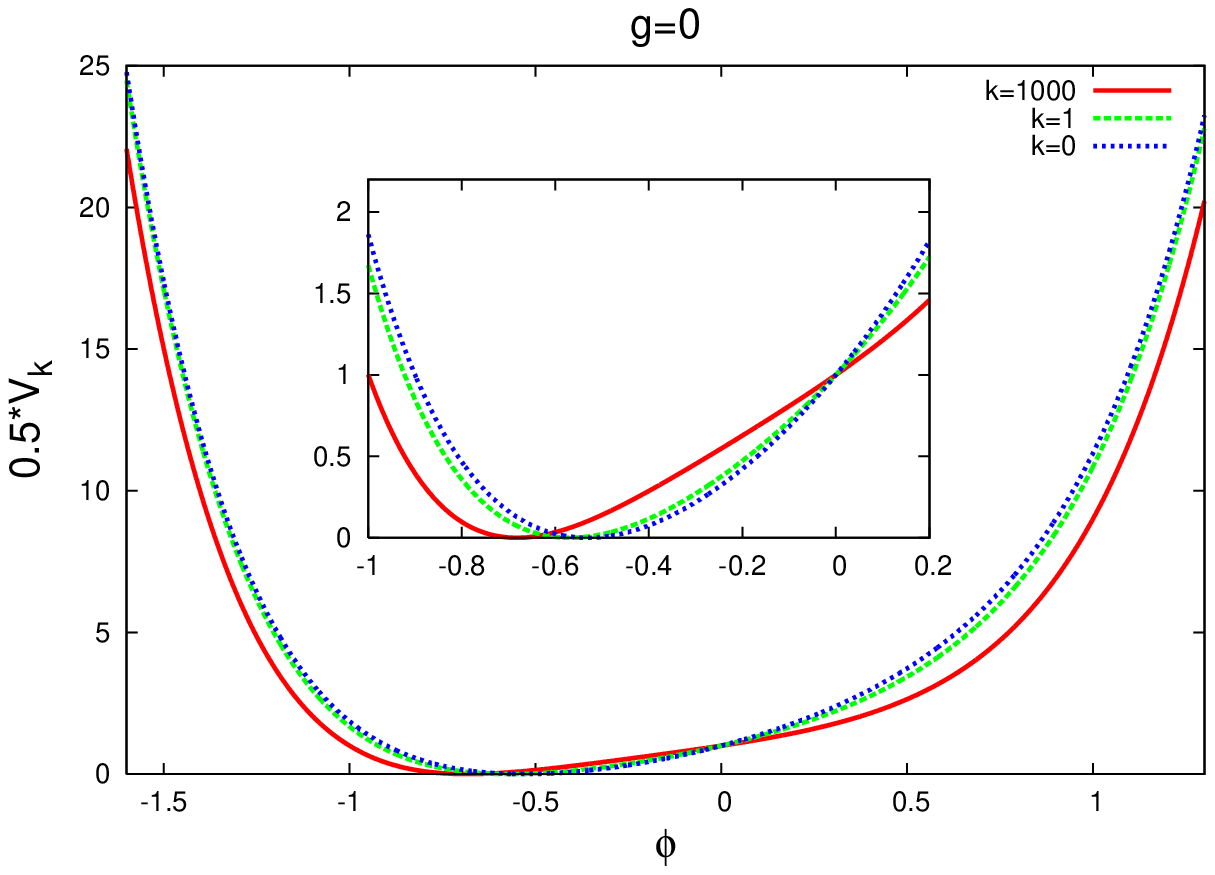}\hfill
\includegraphics[width=.49\textwidth]{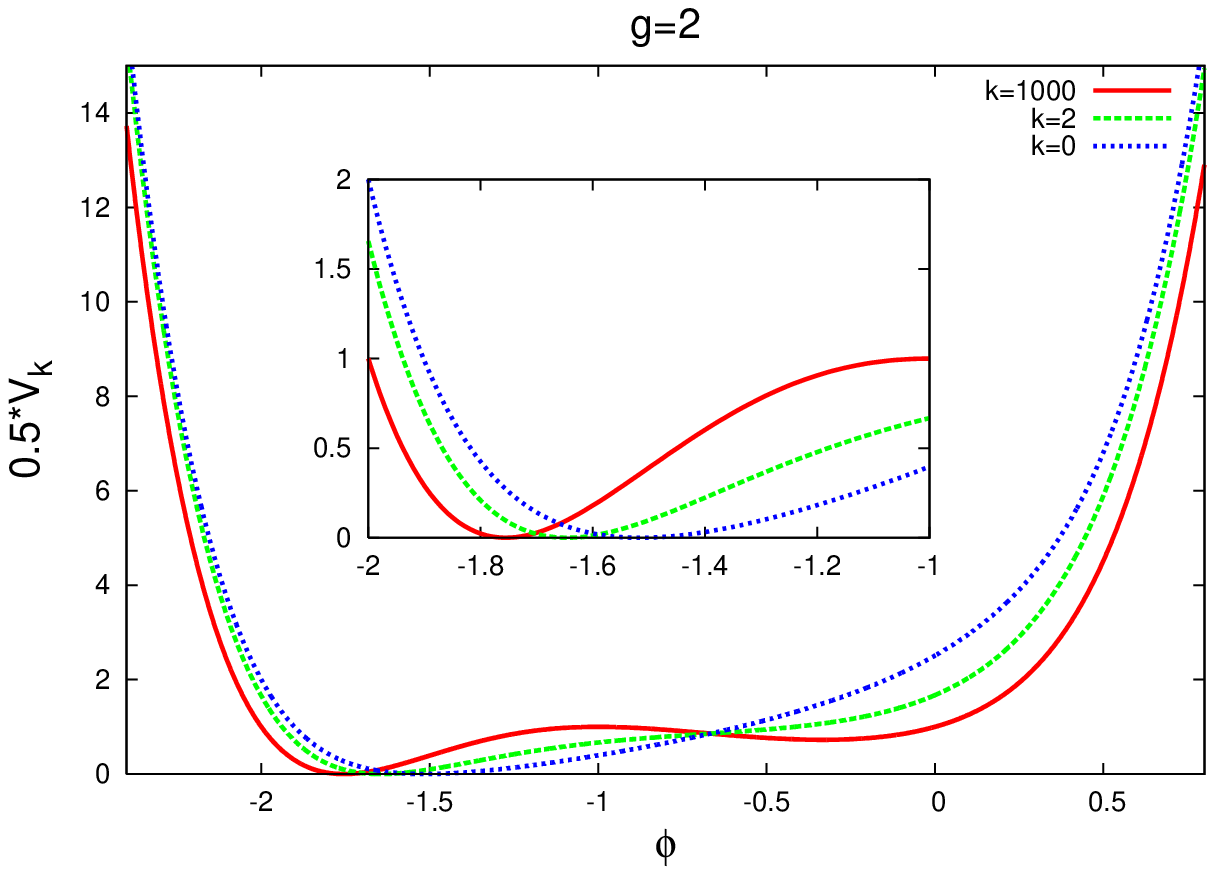}
\caption{
 The effective potential $V_k$ obtained from
 the solution $W_k$ to the
 \emph{partial differential equation}
 \eqref{reflow3} 
 The left panel
 shows $g=0$ and the right panel $g=2$.\label{fig:EffPot}}
}
For convex superpotentials, the solutions obtained from the polynomial
expansions and from solving the partial differential equation are almost
identical. But in the non-convex case, the polynomial expansion fails to
reproduce the correct asymptotic form of the superpotential. Non-convex
classical superpotentials pose a numerical challenge as they might lead to
instabilities originating from the singularity at $W_k''(\phi)=-k$. For such
potentials -- corresponding to a large coupling $g$ -- the flow equation also
does not reproduce the correct gap energies $E_1(g)$; see the
PDE row in Table \ref{tab:FirstExcitedState}.  We shall see that
similar conclusions hold for other regulators in the flow equation.

\subsection{Exponential and $\theta$  regulator}
We want to compare the results obtained with the Callan-Symanzik regulator --
which serves only as an IR regulator but does not suppress the UV --
with an exponential and a $\theta$  regulator,
\begin{equation}
  \begin{aligned}
r^{(\text e)}_1(p^2,k)&=k\cdot e^{-p^2/k^2} &&\text{exponential regulator}\\
r^{(\theta)}_1(p^2,k)&=\sqrt{k^2-p^2}\cdot\theta(k^2-p^2)&&\theta\text{
    regulator.}\label{AltReg1}
   \end{aligned}
 \end{equation}
In contrast to the infrared Callan-Symanzik regulator used in \eqref{reflow3},
these regularize the IR and UV. The corresponding flow
equations for the superpotential read 
\begin{equation}
  \begin{aligned}  
\partial_k W^{(\text e)}_k(\phi)
&=\frac{1}{2k^2}\int\limits_{-\infty}^\infty\frac{dp}{2\pi}
 \frac{(k^2+2p^2)e^{-p^2/k^2}}
{p^2+(W''_k(\phi)+ke^{-p^2/k^2})^2}\\
\partial_kW^{(\theta)}_k(\phi)&=
\frac{1}{4\pi}\frac{k}{\vert k^2-{W''_k}^{\,2}\vert}
\left(\pi\left(1-\hbox{sign}\,W''_k\right)
+2\arctan \frac{\vert k^2-{W''_k}^2\vert}{2kW''_k}\right).\label{AltReg3}
\end{aligned}
\end{equation}
Note, that for the $\theta$ regulator the integral \eqref{reflow1} can be
calculated analytically. The numerical results in Table
\ref{tab:FirstExcitedState1} have been obtained from the solutions to these
partial differential equations.  For the exponential regulator we have taken
$\phi\in(-20,20)$ and the integration over $p$ from $-5k$ to $5k$. For the
$\theta$ regulator,  we have used $\phi\in(-50,50)$.
 \TABLE{
\begin{tabular}{c|cccccccccc}
g&0.0&0.2&0.4&0.6&0.8&1.0&1.2&1.4&1.6&1.8\\ \hline
CS&2.203&2.137&2.062&1.979&1.890&1.798&1.710&1.633&1.584&1.590\\
exp& 2.195&2.130&2.055&1.972&1.884&1.791&1.701&1.622&1.569&1.684\\
$\theta$&2.197&2.132&2.058&1.975&1.888&1.794&1.705&1.626&1.576&1.581\\ \hline
exact&2.022&1.970&1.905&1.827&1.738&1.639&1.534&1.426&1.323&1.235
\end{tabular}
\caption{Energy of the first excited state for the
classical superpotential 
\eqref{polex1} 
with $(e,m,a)=(1,1,1)$ and varying
$g$ calculated from the solution to the partial differential equation 
\eqref {reflow1}
with Callan-Symanzik, exponential and $\theta$ regulators.
\label{tab:FirstExcitedState1}}
}
The results for the three different regulators are depicted in Table
\ref{tab:FirstExcitedState1}. They are almost identical,  but all differ
on the $\sim10\%$ level from the exact values displayed in the last row of the table. 
Higher precision thus requires a next-to-leading order calculation in the
derivative expansion including a wave-function renormalization. 

\section{Wave function renormalization } 
\label{sec:WFR}

To next-to-leading-order in the derivative expansion, a field-dependent wave
function renormalization is included in the truncation,
\begin{align} 
\Gamma_k[\phi,F,{\psib},\psi]&=
\int d\tau\suint
\left[\frac12 \Zf_k(\Phi)  
K\Zf_k(\Phi)+i\cdot
W_k(\Phi)\right]\nonumber\\
 & =
  \int d\tau
  	\left[\frac{1}{2}\Zf'_k(\phi)^2\dot\phi^2-i\Zf'_k(\phi)^2\psib\dot\psi
-i\Zf'_k(\phi)\Zf''_k(\phi)\dot\phi\psib\psi
 +\frac12 \Zf'_k(\phi)^2 F^2\right.\nonumber\\
  &\left.\phantom{\int d\tau
  [\frac12+}-\Zf_k''\Zf_k'F\psib\psi+iFW'_k(\phi)-iW''_k(\phi)\psib\psi\right]
\label{wfr1}
    \end{align}
with a field dependent function $\Zf_k(\phi)$. The operator
$K$ has been defined in \eqref{reg3} and primes denote
derivatives with respect to $\phi$. The results
of the last sections are recovered for $\Zf_k(\Phi)=\Phi$. 

In the spirit of functional optimization \cite{Pawlowski:2005xe}, we choose a
spectrally adjusted regulator \cite{Gies:2002af,Pawlowski:2001df} which
includes the wave function renormalization,
\begin{align}
\Delta S_k=\frac12\int d\tau\suint\; \Zf_k'(\bar\Phi)^2\Phi
\left(ir_1+r_2K\right)\Phi\,, 
\label{wfr2}
\end{align}
where $\Zf_k'$ is evaluated at a background field $\bar\Phi=(\bar\phi,0,0)$. The
value of $\bar\phi$ can be viewed as a parameter labeling a class of regulator
functions. In components, the cutoff action reads
\begin{align}
\Delta S_k=\int d\tau\;\Zf_k'(\bar\phi)^2
\left(\frac12 p^2\phi 
r_2\phi+\frac12 Fr_2F+iFr_1\phi+\psib(pr_2-i r_1)\psi\right)\,.\label{wfr3}
\end{align}
Again, the flow of $\Zf_k$ can be read off from various operators. The
simplest choice is given by the prefactor of the $F^2$ term,
cf. eq.~\eqref{wfr1}, since no time derivatives are involved here. 
After the projection onto the $F^2$ term at vanishing $\psib\psi$ and a constant 
scalar field, we obtain the flow equations for the Callan-Symanzik regulator
\begin{equation}
  \begin{aligned}
\partial_k W'_k(\phi)
=& -W'''_k \frac{\N}{4\cD^2}\\
\Zf'_k(\phi)\partial_k\Zf'_k(\phi)=
&\left(
\frac{4\Zf''_k(\phi)W'''_k(\phi)}{\cD}-\big(\Zf_k''(\phi)\Zf'_k(\phi)\big)'
-\frac{3\Zf_k'(\phi)^2W'''_k(\phi)^2}{4\cD^2}\right)\frac{\N}{4\D^2}\,,
\label{wfr4}
\end{aligned}
\end{equation}
where we have introduced the abbreviations
\begin{align}
 \N=(1+k\partial_k)\Zf'_k(\bar\phi)^2\mtxt{and}
\D=W''(\phi)+k\Zf'_k(\bar\phi)^2\,.
\label{wfr5}
\end{align}
To solve this system of coupled equations, we need to pick a value for the
background field $\bar\phi$. Since we are interested in the excited-state
energy, a reasonable choice would be $\bar\phi=\phi_{\text{min}}$. Since
$\phi_{\text{min}}$ is not a priori known but a result of the flow, this would
require an iterative construction of the RG trajectory. Instead we make a
technically much simpler choice and identify the background field $\bar\phi$
with the fluctuation field $\phi$. Since all functions in the action are
parameters of the background field $\bar\phi$, e.g., $\Zf_k(\phi)\equiv
\Zf_k(\phi,\bar\phi)$, identifying $\bar\phi=\phi$ goes along with an
approximation. This becomes obvious from the fact that, e.g.,
$\Zf_k'(\phi,\bar\phi=\phi)\equiv \partial_\phi
\Zf_k(\phi,\bar\phi)|_{\bar\phi=\phi} \neq \partial_\phi\Zf_k(\phi,\phi)$. By
setting $\bar\phi=\phi$, we ignore this latter difference. This approximation
is well known in the context of background-field flows
\cite{Reuter:1993kw,Reuter:1997gx}, and the resulting flow can be viewed as a
generalized propertime flow \cite{Litim:2002xm,Gies:2002af}. As experience
demonstrates, the error made by this approximation is outweighed by the
improvement arising from the better spectral adjustment of the regulator, see,
e.g., \cite{Bonanno:2000yp}. Our results indeed confirm this conjecture.

Including the wave function renormalization, the on-shell effective bosonic
action at next-to-leading order in the derivative expansion is
\begin{align} 
\Gamma_k[\phi, \psi=0,\bar\psi=0]=\int d\tau \left[ \frac{1}{2}
  \left( \partial_\tau \Zf_k(\phi) \right)^2 + V_k(\phi)\right], \qquad
V_k(\phi) = \frac{1}{2}
  \left(\frac{W'_k(\phi)}{\Zf'_k(\phi)}\right)^2\,.
\label{flowe3}
\end{align}
At $k=0$, the energy gap results from the curvature of the effective potential with
respect to canonically normalized fluctuations $\chi=\Zf(\phi)$, for which we
have the standard kinetic term $\sim( \partial_\tau \chi)^2$. Hence, the
energy of the first excited state for unbroken supersymmetry is
\begin{align}
E_1=\lim_{k\rightarrow  0}\left.\sqrt{\frac{d^2V_k(\Zf_k^{-1}(\chi))}{d\chi^2}
}  \right|_{\chi_{\text{min}}=\Zf(\phi_{\text{min}})} 
 {=}\lim_{k\rightarrow 0}
 \frac{W''(\phi)}{(\Zf'(\phi))^2} \bigg|_{\phi=\phi_{\text{min}}}.
	\label{flowe7} 
\end{align}
In Table \ref{tab:FirstExcitedState2}, the energy gap $E_1(g)$ for
$(e,m,a)=(1,1,1)$ and various couplings $g$ is compared with those obtained
without wave function renormalization. The flow with wave function
renormalization leads to much better results as compared to the flow without
wave function renormalization. The agreement is very satisfactory with errors
on the $\sim 1\%$ level even for couplings of order 1. We conclude that the
flow equation is able to capture nonperturbative physics in supersymmetric
quantum systems with a reasonable precision.

\TABLE{
\begin{tabular}{c|cccccccccc}
g&0.0&0.2&0.4&0.6&0.8&1.0&1.2&1.4&1.6&1.8\\ \hline
PDE&2.203&2.137&2.062&1.979&1.890&1.798&1.710&1.633&1.584&1.590\\
PDE+WF&2.089&2.031&1.961&1.879&1.788&1.690&1.589&1.489&1.402&1.341\\
exact&2.022&1.970& 1.905&1.827&1.738&1.639&1.534&1.426&1.323&1.235
\end{tabular}
\caption{Energy of the first excited state for the
classical superpotential 
\eqref{polex1}
with $(e,m,a)=(1,1,1)$ and varying
$g$ calculated from the solution to flow equations with Callan-Symanzik
regulator without and with wave function renormalization.
\label{tab:FirstExcitedState2}}
}
\section{Summary of the numerical results}
\label{sec:Compare}

\subsection{The energy of the first excited state}

We find that the polynomial approximation and the solution of the partial
differential equation \emph{without} wave function renormalization for convex
superpotentials converge to the same value independent of the regulator, see
Sect. \ref{sect:polyexp}. Depending on the parameters of the classical
superpotential, we obtain an accuracy of 10\% for a small mass parameter
$(m\approx 1)$ and 2\% for larger mass parameters $(m\approx 3)$.  Inclusion
of the wave function renormalization improves the results for the energy gap
considerably. We achieve an accuracy of 3\% for $m=1$.  Due to the presence of
the auxiliary field, the wave function renormalization has contributions of
order $p^0$ in the momentum and the $F^2$-term -- which is neglected without
wave function renormalization -- contributes to the on-shell potential
$V_k(\phi)$. This effect is more pronounced for small mass parameters as the
anomalous dimension scales with the inverse of $m$.  For large $m$, the
anomalous dimension is small so we do not expect large contributions in
agreement with the numerical results.  Figure \ref{fig:CompareEnergy}
summarizes the results for the energies $E_1(g)$ obtained from the different
approximation schemes for $m=1$ and $m=3$.  The explicit values are listed in
Tables \ref{tab:FirstExcitedState}-\ref{tab:FirstExcitedState2}.

\FIGURE[t]{
\includegraphics[width=.49\textwidth]{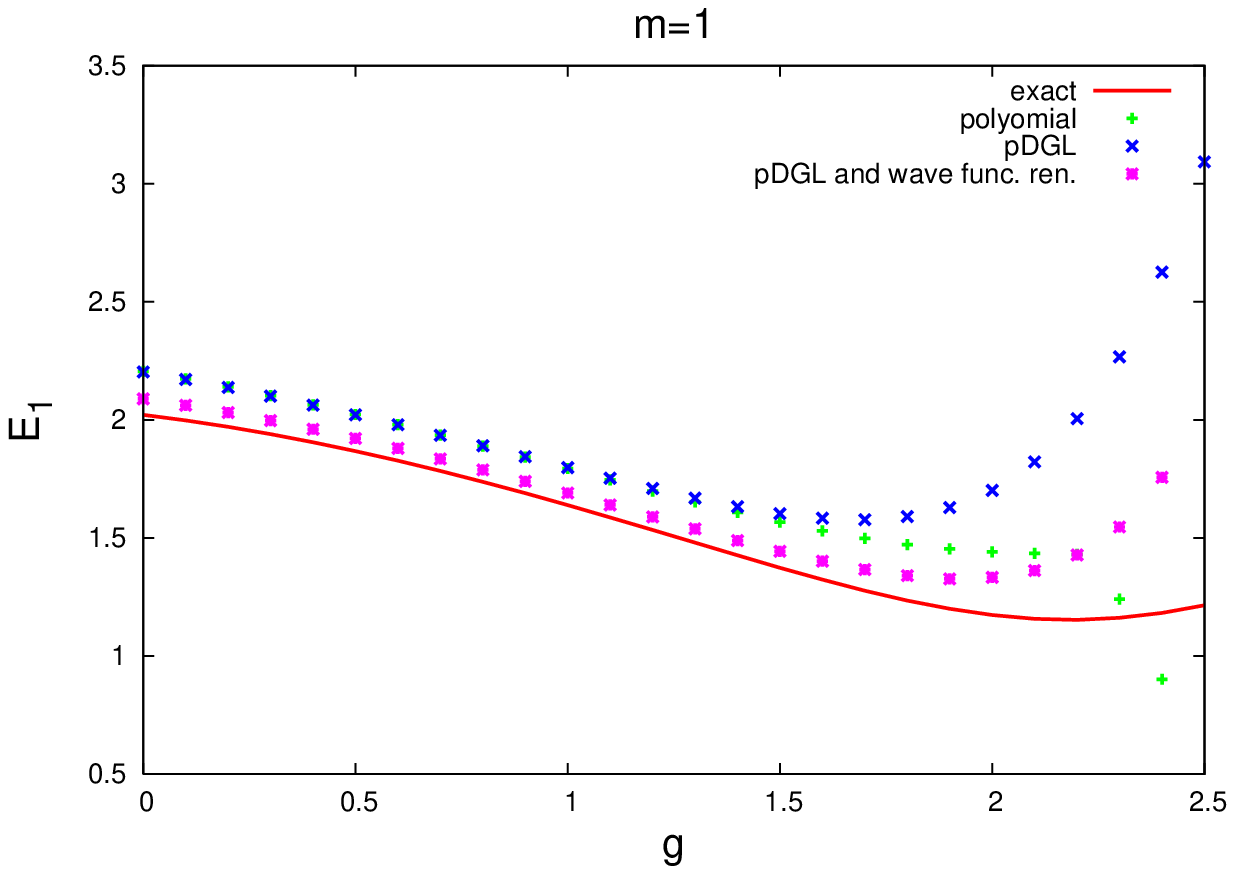}\hfill
\includegraphics[width=.49\textwidth]{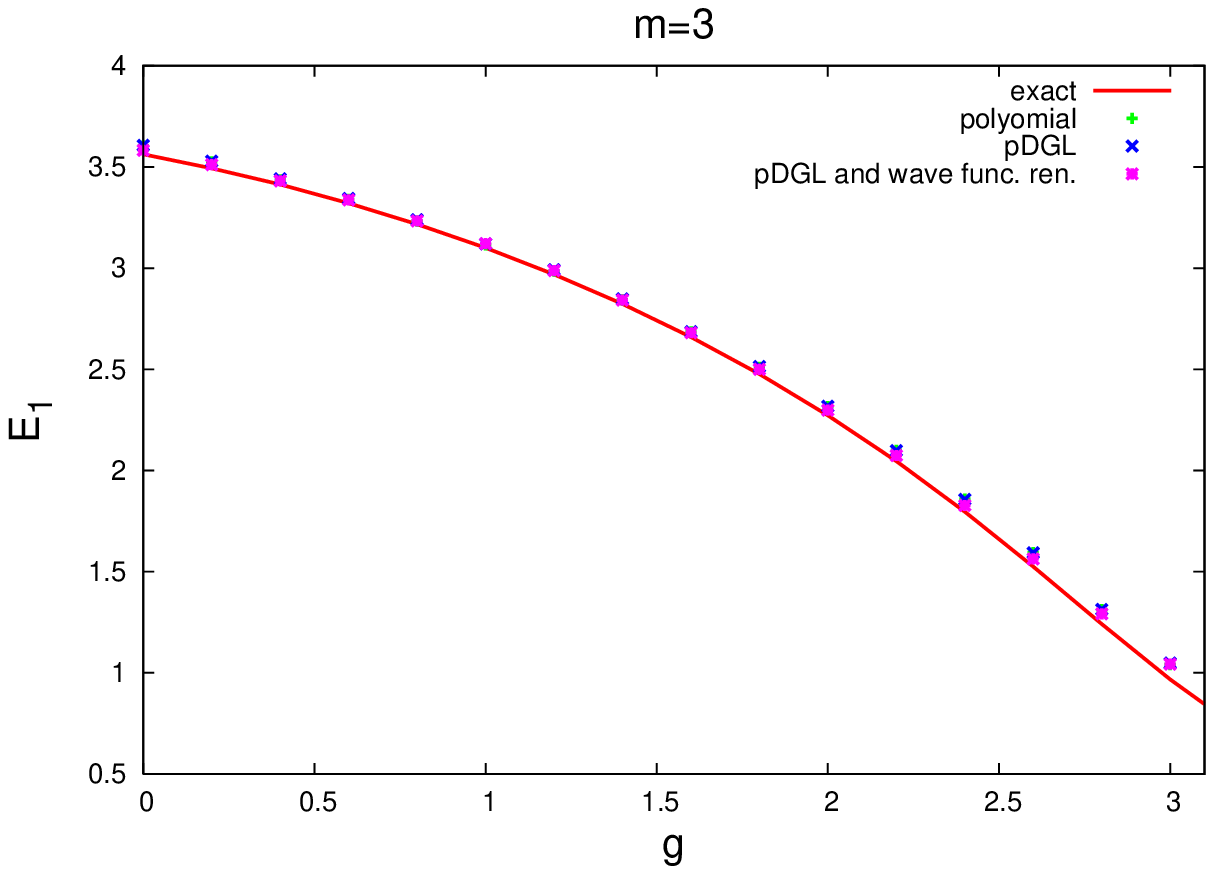}
\caption{We compare the energy gap $E_1(g)$ computed with different
approximation schemes for the classical superpotential
$W_{\rm cl}=1+m\phi+g\phi^2+\phi^3$ with $m=1$ (left panel) and $m=3$ (right
panel). For convex initial potentials, we obtain both a good convergence and
a satisfactory accuracy of the next-to-leading order derivative expansion
including a wave function renormalization. Beyond convex initial potentials,
e.g., for larger couplings $g>\sqrt 3$ for $m=1$, significant deviations from
the exact result are observed, indicating a less controlled convergence
behavior. 
}
\label{fig:CompareEnergy}}

The parameter space of large-$a$ couplings is explored in
Fig.~\ref{fig:CompareEnergyA}. Here, we have used $e=m=g=1$, implying that the
initial potential is always convex. First, we observe that the excited-state
energy from the polynomial expansion converges rapidly to that taken from the
full solution at leading-order. The deviations from the exact result are again
on the $\sim10\%$ level. This is greatly improved at next-to-leading-order
including the wave function renormalization. Here, the results match the exact
values with an error on the 1\% level or below. The agreement holds over the
whole coupling range from the weak- to the deeply nonperturbative
strong-coupling regime.

\FIGURE{
\includegraphics[width=.49\textwidth]{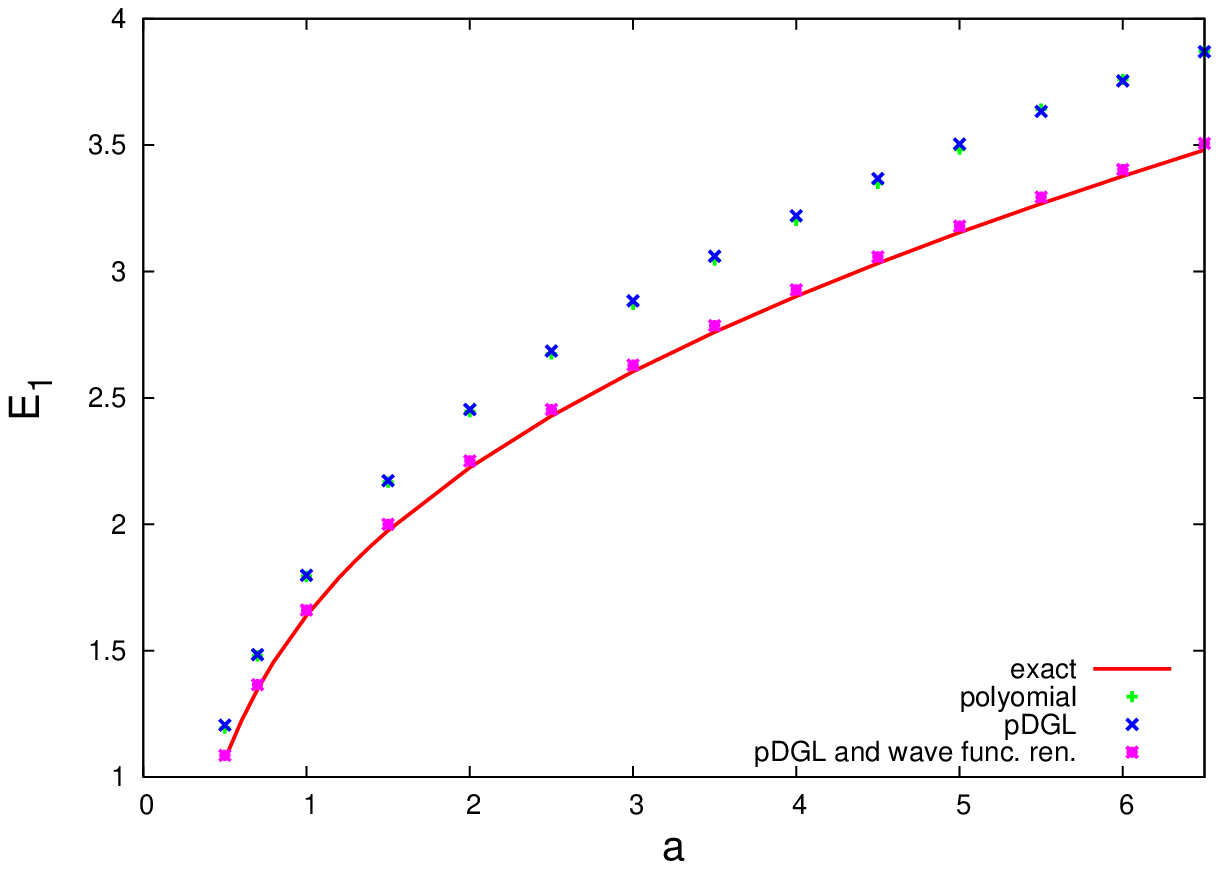}
\caption{Energy gap $E_1(a)$ versus coupling $a$ for $e=m=g=1$ (convex initial
  potentials). We observe a good convergence of the polynomial expansion. At
  next-to-leading-order derivative expansion including a wave function
  renormalization, satisfactory quantitative accuracy is obtained for the
  whole coupling range and even at strong coupling, demonstrating the
  nonperturbative capabilities of the functional RG.  }
\label{fig:CompareEnergyA} 
 }

The overall picture confirms that the functional RG employing the
super-covariant derivative expansion captures the physics of the first excited
state well beyond the perturbative small-coupling regime. For initial boundary
conditions given in terms of classical convex potentials, the derivative
expansion appears to converge well and reaches a very satisfactory accuracy level
already at next-to-leading order.

For combinations of couplings where the initial potential is non-convex, e.g., $g>\sqrt
3$ for $e=m=a=1$, there is clearly room for improvements, as the deviations of
the excited-state energy from the exact result become large. Though the inclusion of
a wave function renormalization at next-to-leading order improves the result
significantly, the accuracy remains poor, see
Fig.~\ref{fig:CompareEnergy}. Moreover, as the next-to-leading-order
correction becomes of the same order as the leading order, the convergence of
the derivative expansion may become questionable. On the other hand, it is
important to note in this context that the hierarchy of the derivative
expansion is interwoven more strongly for the supersymmetric version than for
non-super\-symmetric systems. In the present case, also
next-to-next-to-leading order operators can contribute to the flow of the
superpotential. These contributions may be relevant for non-convex initial
potentials and thus restore the convergence properties of the derivative
expansion.

\subsection{The global structure of the effective potential}

Whereas the polynomial expansion does rather well for the excited-state energy
for the convex case, we observe its break-down beyond this restricted case:
For instance for $g^2>3$ at $e=m=a=1$, the classical superpotential ceases to
be convex. Here, the polynomial approximation fails for asymptotic values of
the field, since it tries to provide for a polynomial solution of the partial
differential equation near the expansion point, where the low-energy effective
potential $V_k$ becomes flat.  The global structure of the effective potential
for $g=2$ calculated from the partial differential equation \eqref{reflow3} and
the polynomial approximation with Callan-Symanzik regulator are plotted in
Figure \ref{fig:GlobalStructurePotential} together with the classical
potential.
\FIGURE{
\includegraphics[width=.49\textwidth]{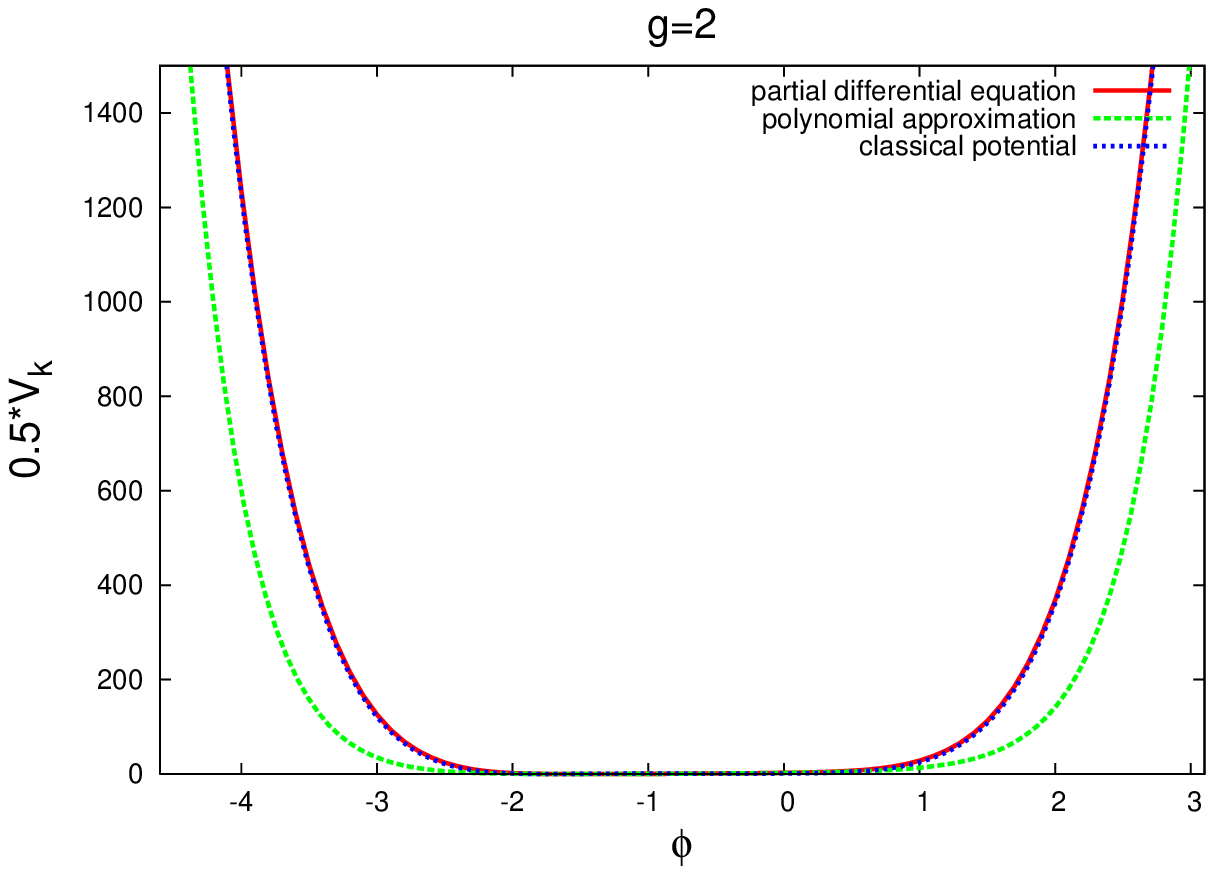}
\caption{The effective potential $W'(\phi)^2$ with the Callan-Symanzik regulator
for a nonconvex $W_{\rm cl}$ with $g=2$. The polynomial expansion fails to
reproduce the global structure of the effective potential, whereas the full
numerical solution of the superpotential flow agrees well with the expectation.}
\label{fig:GlobalStructurePotential}}
As expected the polynomial approximation is not able to reproduce the correct
global structure whereas the partial differential equation is able to do so.
The other regulators lead to the same global structure of the effective 
potential.

\section{Conclusions}
In this paper, we have presented a functional RG approach to supersymmetric
quantum mechanics. Our approach is formulated in terms of an exact and
manifestly supersymmetric flow equation for the effective action which is a
supersymmetric variant of the Wetterich equation. We have used the
supersymmetric off-shell formulation which is the crucial ingredient to
maintain the simple one-loop structure of the flow equation.  The approach can
straightforwardly be generalized to other supersymmetric models based on a
real superfield.

We solve the flow equation nonperturbatively in a systematic and consistent
approximation scheme based on an expansion of the effective action in powers
of field operators of increasing numbers of supercovariant derivatives. To
leading order, this yields a flow equation for the superpotential -- a
supersymmetric analogue of the local-potential approximation; a field
dependent wave function renormalization appears in the flow to next-to-leading
order.

In the present work, we focus on unbroken supersymmetry by considering only
 superpotentials whose highest power is even. As a physical observable, we
 concentrate on the
 energy of the first excited state resulting from the effective potential. A
comparison with the exact solution provides information about the convergence
of the derivative expansion. Our results confirm that the functional RG is
indeed capable of describing the system over the whole range from weak to
strong coupling. Our approach works particularly well for initial convex
potentials. Here, first quantitative estimates can already be obtained from a
simple polynomial expansion of the superpotential. For the excited-state
energy, the polynomial expansion also converges nicely, whereas the solution
of the full partial differential equation for the superpotential is required
for global properties of the potential. Since the excited-state energy is a
physical quantity, it should also be universal in an RG sense. In fact, our
results show little dependence on the regulator which confirms this required
universality. At next-to-leading order, the inclusion of a wave function
renormalization improves the quantitative accuracy considerably. For convex
potentials, the functional RG result agrees with the exact result within an
error on the $\sim 1\%$ level even at strong coupling. 

As soon as the initial potential becomes non-convex, the flow-equation result
for the energy to lowest order starts to deviate significantly from the exact
result. As is already known from standard quantum mechanics, the relevant
tunneling processes are associated also with higher orders in the derivative
expansion. Inclusion of the wave function renormalization indeed improves our
result, even though sizable deviations from the exact result still remain.
The reason for this can be anticipated: supersymmetry forces us to
organize the expansion in powers of the super-covariant derivative. This,
however, mixes different orders of time derivatives; e.g, in the off-shell
version of any supersymmetric theory with a scalar multiplet, the auxiliary
field and the derivative of the scalar field occur on equal footings. This is
visible, for example, in the supersymmetry transformation of $\psi$ being
proportional to $\dot\phi-iF$, see \eqref{susy11}. On the other hand, we expect
that the low-lying excitation energies are mainly determined by the
long-wavelength fluctuations, such that an expansion in time derivatives of
the field should be well justified. 

The crucial observation in this context is that the super-covariant derivative
expansion contains terms without
time derivatives also at higher super-covariant derivative order, for instance,
$\Phi [(D \bar D) \Phi]^2\sim F^3+\dots $. In particular, these $F$-potential
terms can directly contribute to the flow of the superpotential. Since these
terms are generated sizably only at larger values of the coupling, it is
natural to expect that they can exert a pronounced influence on the energy gap
at large coupling. As even higher-order operators will not take a direct influence on
the flow of the superpotential, it is conceivable that the excited-state
energy converges at this next-to-next-to-leading order of the super-covariant
derivative expansion. Otherwise, the convergence and use of this expansion in the
tunneling regime would be questionable. 

A study of these higher orders giving access to operators with higher powers
of $F$ are also needed for the case of broken supersymmetry. In this case, a
nonzero vacuum expectation value of $F$ is expected to occur, the description
of which requires knowledge of the effective potential of this auxiliary
field. 

The models considered here can be obtained by a dimensional reduction from the
$2d$ Wess-Zumino model with $N=1$ supersymmetry.  This in part is the reason
that most structural results of the present work also apply to this
two-dimensional field theory, for example to the form of the cutoff action and
the structure of the flow equations. The super-covariant derivative-expansion
techniques are straightforwardly generalizable.  Work in this direction is
in progress.

\acknowledgments{
Helpful discussions with C.~Wozar and T.~Fischbacher are gratefully
acknowledged. GB acknowledges support by the Evangelisches Studienwerk
and FS by the Studienstiftung des deutschen Volkes. This
work has been supported by the DFG grants Wi 777/8-2 and Gi 328/5-1
(Heisenberg program).}
\appendix


\section{The flow equations in superspace}

In this appendix we sketch the derivation of the flow equation 
for the superpotential in superspace. The equivalence of this  manifestly
supersymmetric derivation with 
the one in component form will be shown afterwards.
The superspace-coordinates $(x,\theta,\thetab)$ are
denoted by $z$.

The supertrace that defines the flow of the effective action translates 
into a superspace integral:
\begin{align}
 \p_k \G_k&=\half \int \,dz\, dz'\, \p_k R_k(z,z') G_k(z',z)\,,\qquad
G_k=(\Gt_k+R_k)^{-1}
\end{align}
As in the component formulation the fields are taken to be
constant to calculate the Green's function $G_k(z',z)$.
In addition the expression is expanded in terms of the covariant 
derivatives $D$ and $\Dbar$. To zeroth order in the covariant 
derivatives one finds 
\begin{multline}
 i\intt  \p_t W(\P)=\half \int\!\frac{dp}{2\pi}\, 
 \tttt\left(i\p_t r_1(p)+\p_t r_2(p)K(p)\right)\times\\\times
 \deltatt \frac{hK(p)-i\cW''(\P)}{hp^2+(\cW''(\P))^2} \deltatt\, .
\end{multline}
Note that in momentum space the operator $K=\frac12(D\Dbar-\Dbar D)$  
still contains derivatives with respect to the Grassmann-coordinates. 
These derivatives act on the first entry of the adjacent delta-functions. 
The only two contributions that remain after an integration over $\t'$ and $\tb'$ are the ones where the highest Grassmann derivative acts on one and only one of the delta functions inside the integral. Therefore we get
\begin{align}
 \intt \p_t W(\P)=
\half \int\!\frac{dp}{2\pi} \,d\t d\tb \,
\left(\frac{h\p_t r_1(p)-\cW''(\Phi)\p_t r_2(p)}
{hp^2+(\cW''(\P))^2}\right) .\label{ap3}
\end{align}
For the lowest component of the superfield this is exactly the 
flow equation \eqref{flow29}.

To prove the equivalence of this derivation to the 
one in given in the main body of the paper 
we observe that the transition from component 
to superfield formulation can be achieved
with the linear operators \mbox{$P(\t,\tb)=(1,\tb\t,-\t,\tb)$}  and
$P^T(\t,\tb)=(1,\tb\t,\t,-\tb):$
\begin{align}
 \P&=P_i(\t,\tb) (\phi,F,\psib,\psi)_i
 =(1,\tb\t,-\t,\tb)_i (\phi,F,\psib,\psi)_i
 =\phi+\tb\psi+\psib\t+\tb\t F\nonumber
\\&= (\phi,F,\psib,\psi)_i(1,\tb\t,\t,-\tb)_i=(\phi,F,\psib,\psi)_iP^T_i(\t,\tb)\, .
\end{align}
In the other direction the operator $Q(\t,\tb)=(\tb\t,1,-\tb,-\t)$ must be applied:
\begin{align}
 (\phi,F,\psib,\psi)_i&=\intt Q_i(\t,\tb)\P =\intt (\tb\t,1,-\tb,-\t)_i\, 
(\phi+\tb\psi+\psib\t+\t\tb F)\,.
\end{align}
Note that as expected $P_i(\t',\tb')Q_i(\t,\tb)=Q_i(\t',\tb')P^T_i(\t,\tb)=\deltatt$ and $\intt Q_i(\t,\tb)P_j(\t,\tb)=\intt P^T_i(\t,\tb)Q_j(\t,\tb)=\delta_{ij}$.
The operator $R_k$ and its inverse can be easily translated from component to superspace formulation using these operators:
\begin{align}
(R_k(x,x'))_{ij}&=\inttt P^T_i(\t,\tb) R_k(z,z')P_j(\t',\tb') \\
 (R_k(x,x'))^{-1}_{ij}&=\inttt Q_i(\t,\tb)(R_k(z,z'))^{-1}Q_j(\t',\tb') \, ,
\end{align}
with $\int dz' R_k(z,z')(R_k(z',z))^{-1}=\delta(z-z')$.
So the flow equations translate into 
\begin{align}
 \int& dx dx'(-1)^{\veps_i}\p_t R_k(x,x')_{ij}G_k(x',x)_{ji}\nonumber\\
 &=\inttt dx dz''(-1)^{\veps_i}Q_i(\t,\tb)P_i(\t',\tb')(\p_t R_k)(x,\t,\tb;z'')G_k(z'';x,\t',\tb')
 	\nonumber\\
 &=\int dz dz''(\p_t R_k)(z,z'')G_k(z'';z)
\end{align}
with $(-1)^{\veps_i}$ $-1$ if $i$ is a fermionic index and $1$ otherwise,
since $(-1)^{\veps_i}P_i(\t,\tb)=P^T_i(\t,\tb)$.

\label{FlowSuperspace}


\section{Initial conditions}
\label{appB}

Throughout this work, we have set the regulator component $r_2=0$. With regard
to the regulator structure \eqref{reg7}, one may wonder whether this choice is
compatible with a sufficient UV suppression of all modes. If not, the initial
condition of the flow would not necessarily coincide with the microscopic
(classical) action, but a separate UV renormalization would be
necessary. 

Indeed, it is easy to see that diagrams containing closed $F$ loops with a
momentum-independent free propagator can give rise to UV divergencies
signaling this insufficient UV suppression.  On the other hand, closed $F$
loops do simply not contribute to the present truncation; this would require,
e.g., the occurrence of $F$ self-interactions $\sim F^3$ which are generated
only at higher-order in the super-covariant derivative
expansion. Perturbatively, they occur at the two-loop level. 
We conclude that there is no danger from $F$ loops up to next-to-leading order
in the derivative expansion.

Indeed, sufficient UV suppression can directly be verified. For large $k$, the
cutoff action $\Delta S_k$ dominates the action in the defining Euclidean path
integral which is of the form \cite{Wetterich:1992yh}
\begin{align}
	e^{-\Gamma_k[\phi,F,\psi,\psib]}=\int\cD\varphi\cD
 \FTild\cD\chi\cD\chib
 e^{-S[\phi+\varphi,F+\FTild,\psi+\chi,\psib+\chib]}
 e^{\frac{\delta\Gamma_k}{\delta\phi}\varphi
 +\frac{\delta\Gamma_k}{\delta F}\FTild
 +\frac{\delta\Gamma_k}{\delta\psi}\chi
 +\frac{\delta\Gamma_k}{\delta\psib}\chib
 -\varphi r_1\FTild-\chi r_1\chib
 }\,.
\end{align}
The integral becomes dominated by small fluctuations around the
classical solutions in the presence of the cutoff. A good estimate is thus
provided by a saddle-point approximation of the path integral. Using the
simple Callan-Symanzik regulator $r_1=k$ as an example, one-loop corrections
are given by
\begin{align*}
	\Gamma_{k,1\text{loop}}=-\frac12\int \limits_{-\infty}^\infty\frac{dp}{2\pi}
	\ln\frac{S_{\phi\phi}S_{FF}-S_{F\phi}^2}{S_{\psi\psib}S_{\psib\psi}}
	=-\frac12\int\limits_{-\infty}^\infty
        \frac{dp}{2\pi}\ln\left(1+\frac{iFW'''}{p^2+(W''+k)^2}\right).
\end{align*}
Rescaling $p$ with $k$  yields 
\begin{align*}
	\Gamma_{k,1\text{loop}}=-k\int\limits_{-\infty}^\infty \frac{d\tilde 
	p}{2\pi}\ln\left(1+\frac1{k^2}\frac{iFW'''}{\tilde p^2+(W''/k+1)^2}\right).
\end{align*}
This integral vanishes for $k\to\infty$ so that no UV counterterms are
necessary to define the initial conditions. The starting point of the
flow equation is indeed the classical action.


\begin{thebibliography}{10}

\bibitem{Feo:2004kx}
A.~Feo, {\it {Predictions and recent results in SUSY on the lattice}},  {\em
  Mod. Phys. Lett.} {\bf A19} (2004) 2387--2402,
  [\href{http://arxiv.org/abs/hep-lat/0410012}{{\tt hep-lat/0410012}}].

\bibitem{Giedt:2006pd}
J.~Giedt, {\it {Deconstruction and other approaches to supersymmetric lattice
  field theories}},  {\em Int. J. Mod. Phys.} {\bf A21} (2006) 3039--3094,
  [\href{http://arxiv.org/abs/hep-lat/0602007}{{\tt hep-lat/0602007}}].

\bibitem{Bergner:2007pu}
G.~Bergner, T.~Kaestner, S.~Uhlmann, and A.~Wipf, {\it {Low-dimensional
  supersymmetric lattice models}},  {\em Annals Phys.} {\bf 323} (2008)
  946--988, [\href{http://arxiv.org/abs/0705.2212}{{\tt arXiv:0705.2212}}].

\bibitem{Kastner:2008zc}
T.~Kastner, G.~Bergner, S.~Uhlmann, A.~Wipf, and C.~Wozar, {\it
  {Two-Dimensional Wess-Zumino Models at Intermediate Couplings}},  {\em Phys.~Rev.} {\bf D78} (2008)
  095001,  [\href{http://arxiv.org/abs/0807.1905}{{\tt arXiv:0807.1905}}].

\bibitem{Aoki:2000wm}
K.~Aoki, {\it {Introduction to the nonperturbative renormalization group and
  its recent applications}},  {\em Int. J. Mod. Phys.} {\bf B14} (2000)
  1249--1326.

\bibitem{Berges:2000ew}
J.~Berges, N.~Tetradis, and C.~Wetterich, {\it Non-perturbative renormalization
  flow in quantum field theory and statistical physics},  {\em Phys. Rept.}
  {\bf 363} (2002) 223--386. hep-ph/0005122.
  
\bibitem{Litim:1998nf}
D.~F. Litim and J.~M. Pawlowski, {\it {On gauge invariant Wilsonian flows}},
  \href{http://arxiv.org/abs/hep-th/9901063}{{\tt hep-th/9901063}}.

\bibitem{Pawlowski:2005xe}
J.~M. Pawlowski, {\it {Aspects of the functional renormalisation group}},  {\em
  Annals Phys.} {\bf 322} (2007) 2831--2915,
  [\href{http://arxiv.org/abs/hep-th/0512261}{{\tt hep-th/0512261}}].

\bibitem{Gies:2006wv}
H.~Gies, {\it {Introduction to the functional RG and applications to gauge
  theories}},  \href{http://arxiv.org/abs/hep-ph/0611146}{{\tt
  hep-ph/0611146}}.

\bibitem{Sonoda:2007av}
H.~Sonoda, {\it {The Exact Renormalization Group -- renormalization theory
  revisited --}},  \href{http://arxiv.org/abs/0710.1662}{{\tt
  arXiv:0710.1662}}.

\bibitem{Horikoshi:1998sw}
A.~Horikoshi, K.-I. Aoki, M.-a. Taniguchi, and H.~Terao, {\it {Non-perturbative
  renormalization group and quantum tunnelling}},
  \href{http://arxiv.org/abs/hep-th/9812050}{{\tt hep-th/9812050}}.

\bibitem{Kapoyannis:2000sp}
A.~S. Kapoyannis and N.~Tetradis, {\it {Quantum-mechanical tunnelling and the
  renormalization group}},  {\em Phys. Lett.} {\bf A276} (2000) 225--232,
  [\href{http://arxiv.org/abs/hep-th/0010180}{{\tt hep-th/0010180}}].

\bibitem{Zappala:2001nv}
D.~Zappala, {\it {Improving the Renormalization Group approach to the
  quantum-mechanical double well potential}},  {\em Phys. Lett.} {\bf A290}
  (2001) 35--40, [\href{http://arxiv.org/abs/quant-ph/0108019}{{\tt
  quant-ph/0108019}}].

\bibitem{Weyrauch:2006}
M.~Weyrauch, {\it Functional renormalization group: Truncation schemes and
  quantum tunneling},  {\em Journal of Molecular Liquids} {\bf 127} (2006)
  21--27. International Conference on Physics of Liquid Matter: Modern
  Problems.

\bibitem{Weyrauch:2006aj}
M.~Weyrauch, {\it {Functional renormalization group and quantum tunnelling}},
  {\em J. Phys.} {\bf A39} (2006) 649--666.

\bibitem{Witten:1981nf}
E.~Witten, {\it {Dynamical Breaking of Supersymmetry}},  {\em Nucl. Phys.} {\bf
  B188} (1981) 513.

\bibitem{Bonini:1993sj}
M.~Bonini, M.~D'Attanasio, and G.~Marchesini, {\it {Renormalization group flow
  for SU(2) Yang-Mills theory and gauge invariance}},  {\em Nucl. Phys.} {\bf
  B421} (1994) 429--455, [\href{http://arxiv.org/abs/hep-th/9312114}{{\tt
  hep-th/9312114}}].

\bibitem{Ellwanger:1994iz}
U.~Ellwanger, {\it {Flow equations and BRS invariance for Yang-Mills
  theories}},  {\em Phys. Lett.} {\bf B335} (1994) 364--370,
  [\href{http://arxiv.org/abs/hep-th/9402077}{{\tt hep-th/9402077}}].

\bibitem{Ginsparg:1981bj}
P.~H. Ginsparg and K.~G. Wilson, {\it {A remnant of chiral symmetry on the
  lattice}},  {\em Phys. Rev.} {\bf D25} (1982) 2649.

\bibitem{Igarashi:2002ba}
Y.~Igarashi, H.~So, and N.~Ukita, {\it {Ginsparg-Wilson relation and lattice
  chiral symmetry in fermionic interacting theories}},  {\em Phys. Lett.} {\bf
  B535} (2002) 363--370, [\href{http://arxiv.org/abs/hep-lat/0203019}{{\tt
  hep-lat/0203019}}].

\bibitem{Bergner:2008ws}
G.~Bergner, F.~Bruckmann, and J.~M. Pawlowski, {\it {Generalising the
  Ginsparg-Wilson relation: Lattice supersymmetry from blocking
  transformations}},  \href{http://arxiv.org/abs/0807.1110}{{\tt
  arXiv:0807.1110}}.

\bibitem{Vian:1998kv}
F.~Vian, {\it Supersymmetric gauge theories in the exact renormalization group
  approach},  \href{http://arxiv.org/abs/hep-th/9811055}{{\tt hep-th/9811055}}.

\bibitem{Bonini:1998ec}
M.~Bonini and F.~Vian, {\it Wilson renormalization group for supersymmetric
  gauge theories and gauge anomalies},  {\em Nucl. Phys.} {\bf B532} (1998)
  473--497, [\href{http://arxiv.org/abs/hep-th/9802196}{{\tt hep-th/9802196}}].

\bibitem{Falkenberg:1998bg}
S.~Falkenberg and B.~Geyer, {\it {Effective average action in N = 1
  super-Yang-Mills theory}},  {\em Phys. Rev.} {\bf D58} (1998) 085004,
  [\href{http://arxiv.org/abs/hep-th/9802113}{{\tt hep-th/9802113}}].

\bibitem{Arnone:2004ey}
S.~Arnone and K.~Yoshida, {\it {Application of exact renormalization group
  techniques to the non-perturbative study of supersymmetric field theory}},
  {\em Int. J. Mod. Phys.} {\bf B18} (2004) 469--478.

\bibitem{Arnone:2004ek}
S.~Arnone, F.~Guerrieri, and K.~Yoshida, {\it {N = 1* model and glueball
  superpotential from renormalization group improved perturbation theory}},
  {\em JHEP} {\bf 05} (2004) 031,
  [\href{http://arxiv.org/abs/hep-th/0402035}{{\tt hep-th/0402035}}].

\bibitem{Rosten:2008ih}
O.~J. Rosten, {\it {On the Renormalization of Theories of a Scalar Chiral
  Superfield}},  \href{http://arxiv.org/abs/0808.2150}{{\tt arXiv:0808.2150}}.

\bibitem{Sonoda:2008dz}
H.~Sonoda and K.~Ulker, {\it {Construction of a Wilson action for the
  Wess-Zumino model}},  \href{http://arxiv.org/abs/0804.1072}{{\tt
  arXiv:0804.1072}}.

\bibitem{Wetterich:1992yh}
C.~Wetterich, {\it {Exact evolution equation for the effective potential}},
  {\em Phys. Lett.} {\bf B301} (1993) 90--94.


\bibitem{Morris:2005ck}
  T.~R.~Morris,
  {\it {Equivalence of local potential approximations}},
  {\em JHEP} {\bf 0507}, 027 (2005),
  [\href{http://arxiv.org/abs/hep-th/0503161}{{\tt hep-th/0503161}}].

\bibitem{Litim:2000ci}
  D.~F.~Litim,
  {\it {Optimisation of the exact renormalisation group}},
  {\em Phys.\ Lett.}   {\bf B486}, 92 (2000)
  [arXiv:hep-th/0005245].
  
\bibitem{Litim:2001up}
D.~F. Litim, {\it Optimised renormalisation group flows},  {\em Phys. Rev.}
  {\bf D64} (2001) 105007, [\href{http://arxiv.org/abs/hep-th/0103195}{{\tt
  hep-th/0103195}}].

\bibitem{Gies:2002af}
H.~Gies, {\it {Running coupling in Yang-Mills theory: A flow equation study}},
  {\em Phys. Rev.} {\bf D66} (2002) 025006,
  [\href{http://arxiv.org/abs/hep-th/0202207}{{\tt hep-th/0202207}}].
  

\bibitem{Pawlowski:2001df}
  J.~M.~Pawlowski,
{\it {On Wilsonian flows in gauge theories}},
  {\em Int.\ J.\ Mod.\ Phys.}   {\bf A16}, 2105 (2001).

\bibitem{Reuter:1993kw}
M.~Reuter and C.~Wetterich, {\it {Effective average action for gauge theories
  and exact evolution equations}},  {\em Nucl. Phys.} {\bf B417} (1994)
  181--214.

\bibitem{Reuter:1997gx}
M.~Reuter and C.~Wetterich, {\it {Gluon condensation in nonperturbative flow
  equations}},  {\em Phys. Rev.} {\bf D56} (1997) 7893--7916,
  [\href{http://arxiv.org/abs/hep-th/9708051}{{\tt hep-th/9708051}}].

\bibitem{Litim:2002xm}
D.~F. Litim and J.~M. Pawlowski, {\it {Completeness and consistency of
  renormalisation group flows}},  {\em Phys. Rev.} {\bf D66} (2002) 025030,
  [\href{http://arxiv.org/abs/hep-th/0202188}{{\tt hep-th/0202188}}].

\bibitem{Bonanno:2000yp}
A.~Bonanno and D.~Zappala, {\it {Towards an accurate determination of the
  critical exponents with the renormalization group flow equations}},  {\em
  Phys. Lett.} {\bf B504} (2001) 181--187,
  [\href{http://arxiv.org/abs/hep-th/0010095}{{\tt hep-th/0010095}}].

\end{thebibliography}

\providecommand{\href}[2]{#2}\begingroup\raggedright\endgroup

\end{document}